\documentclass[twoside,9pt]{article}
\usepackage{extsizes}
\usepackage[super,sort&compress,comma]{natbib} 
\usepackage[version=3]{mhchem}
\usepackage[left=1.5cm, right=1.5cm, top=1.785cm, bottom=2.0cm]{geometry}
\usepackage{balance}
\usepackage{mathptmx}
\usepackage{tcolorbox}
\usepackage{etoolbox}
\usepackage{sectsty}
\usepackage{graphicx} 
\usepackage{lastpage}
\usepackage[format=plain,justification=justified,singlelinecheck=false,font={stretch=1.125,small,sf},labelfont=bf,labelsep=space]{caption}
\usepackage{float}
\usepackage{fancyhdr}
\usepackage{fnpos}
\usepackage[english]{babel}
\addto{\captionsenglish}{%
  
}
\usepackage{array}
\usepackage{droidsans}
\usepackage{charter}
\usepackage[T1]{fontenc}
\usepackage[usenames,dvipsnames]{xcolor}
\usepackage{setspace}
\usepackage[compact]{titlesec}
\usepackage{hyperref}
\usepackage{multirow}
\usepackage{mathtools}
\usepackage{siunitx}
\usepackage{braket} % Braket
\usepackage[sep=0em, offset=1.6em]{simpler-wick}  % Wick contractions

\usepackage{mathrsfs} % For \mathscr
\DeclareMathAlphabet{\mathcal}{OMS}{cmsy}{m}{n} % For \mathcal to be different from \mathscr
\usepackage{epstopdf}%This line makes .eps figures into .pdf - please comment out if not required.
\definecolor{cream}{RGB}{222,217,201}

\DeclareMathOperator{\spn}{span}
\newcommand{\D}{\mathrm{d}}
\newcommand{\gnocchi}{\textsc{GNOCCHI}}
\newcommand{\qsymsq}{\textsc{QSym\textsuperscript{2}}}
\titleformat{\section}          % Format for \section
  {\bfseries\fontsize{12pt}{14pt}\selectfont} % Bold, 12pt font
  {\thesection}                % Section number
  {1em}                        % Space between number and title
  {}                           % Before the title
\begin{document}

%%%TITLE AND AUTHORS%%%
%\title{Title} %The title of the article should be inputed there
%\author{Author Full Name,*a Author Full Name b and Author Full Name c}
%\date{} %do not delete or the date will appear in the generated pdf
%\maketitle
\begin{flushleft}
{\fontsize{18pt}{20pt}\selectfont\textbf{Symmetry-adapted generalised normal-ordered coupled-cluster theory for excited states}}
\end{flushleft}
{\fontsize{12pt}{14pt}\selectfont Nicholas Lee,\textsuperscript{*a} David P. Tew,\textsuperscript{a} and Bang C. Huynh\textsuperscript{*a}}\\\\
Received 00th January 20xx, Accepted 00th January 20xx
DOI: 10.1039/x0xx00000x\vspace{0.5cm}
%%%END OF TITLE AND AUTHORS%%%

%%%FOOTNOTES%%%
\makeatletter
\def\blfootnote{\gdef\@thefnmark{}\@footnotetext}
\makeatother
\blfootnote{\textsuperscript{a}~Physical and Theoretical Chemistry Laboratory, Department of Chemistry, University of Oxford, Oxford OX1 3QZ, United Kingdom}
\blfootnote{\textsuperscript{*}~Corresponding emails: nicholas.lee@chem.ox.ac.uk; bang.huynh@chem.ox.ac.uk}

%%%END OF FOOTNOTES%%%

\begin{abstract}
  Ground and excited electronic states in highly symmetric systems typically possess high degrees of spatial degeneracy as a consequence of point-group symmetry.
  However, many current quantum-chemical methods struggle to accurately describe the strong correlation effects inherently present in these states, thereby precluding the ability to obtain meaningful insights into the electronic structure of the underlying systems.
  Consequently, many of their important chemical and spectroscopic properties cannot be reliably computed and predicted.
  In this article, a new theoretical framework is described that unifies the symbolic treatment of non-Abelian symmetry in \qsymsq{} and the recently developed state-specific multi-reference coupled cluster theory termed \textit{Generalised Normal Ordered Coupled Cluster (GNOCC)} to describe such difficult ground and excited states in a balanced and targeted manner.
  This is ensured by the ability of \qsymsq{} to exploit symmetry orbits to restore any broken spatial symmetries and generate symmetry-adapted multi-determinantal wavefunctions, as well as the ability of GNOCC to dynamically correlate arbitrary spin eigenfunctions in a size-extensive and spin-free manner.
  To illustrate the capabilities of this framework, several ground and excited states in three model systems are examined in detail: (i) octahedral (H\textsubscript{6})\textsuperscript{2+}, (ii) octahedral H\textsubscript{6}, and (iii) tetrahedral Li\textsubscript{4}.
  The results demonstrate that the proposed method can target both degenerate and non-degenerate states, while delivering improved numerical performance relative to conventional single-reference coupled-cluster approaches.

  \begin{figure}[h!]
    \centering
    \includegraphics[width=8cm]{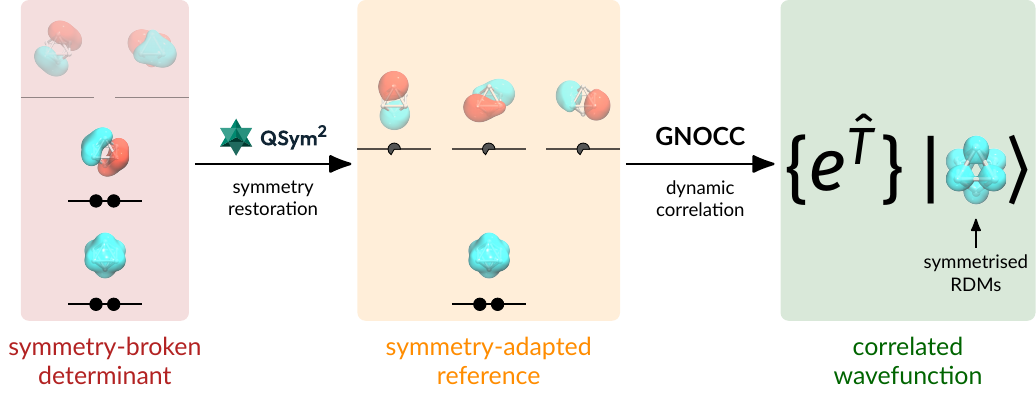}
  \end{figure}
\end{abstract}

\section{Introduction}
  Highly symmetric open-shell systems such as transition-metal complexes and molecular clusters invariably possess strongly correlated electronic ground and excited states as a direct consequence of their high degrees of degeneracy.\cite{article:Tanabe1954a,article:Tanabe1954b,article:Tanabe1956,book:Griffith1961,book:Sugano1970,article:Jackson1993,article:Li1993}
  Unfortunately, the ability to describe these states accurately remains a significant challenge for modern electronic-structure methods.\cite{article:Burke2005,article:Pierloot2017,article:Phung2018,article:Huynh2020}
  This is chiefly because a reliable characterisation of the electronic structure of these states requires a balanced and targeted treatment of both static and dynamical correlation effects.\cite{article:Tew2007,article:Cramer2009,article:Tsuchimochi2016}
  
  To capture the dominant static correlation effects while avoiding unphysical artefacts associated with symmetry-breaking, we exploit symmetry orbits\cite{book:Smith2002} within the representation-theoretic framework of \qsymsq{}, a program for quantum symbolic symmetry.\cite{article:Huynh2024}
  This approach enables the systematic construction of symmetry-adapted multi-determinantal wavefunctions from symmetry-broken determinants,\cite{article:Huynh2020} particularly for systems possessing non-Abelian symmetries.
  By enforcing symmetry adaptation via either projection operators\cite{book:Aktins2011} or non-orthogonal configuration interaction (NOCI),\cite{article:Thom2009b,article:Sundstrom2014b,article:Huynh2020,leeLocalizedSpinRotations2022} the resulting wavefunctions can be unambiguously assigned to both degenerate and non-degenerate multi-reference ground and excited states. 
  This thus provides a robust, symmetry-consistent, and physically meaningful foundation for the subsequent incorporation of dynamical correlation.
  
  To achieve quantitative accuracy, it is further required that dynamical correlation effects be incorporated on top of the multi-determinantal reference wavefunctions. 
  While a wide range of multi-reference coupled cluster (MRCC) theories\cite{lyakhMultireferenceNatureChemistry2012,carskyRecentProgressCoupled,kohnStatespecificMultireferenceCoupledcluster2013,evangelistaPerspectiveMultireferenceCoupled2018} have been developed for this purpose, there has been no clear consensus in the literature regarding the most optimal approach.
  Recently, two of the authors have introduced a state-specific MRCC approach termed \textit{Generalised Normal-Ordered Coupled Cluster (GNOCC)},\cite{leeSpinfreeGeneralizedNormal2026,gunasekeraMultireferenceCoupledCluster2024} which is based on Lindgren's generalised normal-ordered exponential \textit{ansatz}.\cite{lindgrenCoupledclusterApproachManybody1978}
  GNOCC is both spin-free and size-extensive,\cite{leeSpinfreeGeneralizedNormal2026} and extends conventional single-reference coupled cluster theory to correlate arbitrary spin eigenfunctions, as well as both ground and excited states, in a state-specific and chemically balanced manner.
  Within this formalism, dynamical correlation can be systematically incorporated on top of the symmetry-adapted multi-determinantal wavefunctions constructed using \qsymsq{}.
  
  Our aim is to establish a unified framework that combines the theoretical formalisms of \qsymsq{} and GNOCC to recover both \textit{static} and \textit{dynamical} correlation in degenerate electronic states.
  This enables a targeted and accurate description of specific ground and excited states in highly symmetric structures.
  In this article, we present a method that utilises symmetry adaptation on symmetry-broken Hartree--Fock (HF) determinants to recover static correlation arising from degeneracy, and subsequently incorporates dynamical correlation through GNOCC.
  While GNOCC formally admits any spin eigenfunction as a reference state, its practical application has thus far been largely restricted to complete-active-space (CAS) wavefunctions.
  The proposed method therefore constitutes a pilot implementation to use orbit-based multi-determinantal wavefunctions constructed using \qsymsq{} within the GNOCC framework.
  This greatly broadens the scope and applicability of GNOCC as a MRCC method.
  
  The article is organised as follows.
  In Section~\ref{sec:theory}, we first review the principles underlying the spatial symmetry adaptation of multi-determinantal wavefunctions. 
  We then recapitulate the key features of the GNOCC method, including its working equations and the treatment of redundancies in the excitation space. 
  This Section concludes with a procedure for approximating multi-determinantal wavefunctions constructed with \qsymsq{} as CAS-type references compatible with the GNOCC framework.
  Section~\ref{sec:compdetails} outlines the computational details, and Section~\ref{sec:results} presents results for several highly symmetric systems, with particular emphasis on the performance of our proposed method for degenerate electronic states.

\section{Theory}
  \label{sec:theory}
  
  \subsection{Spatial-symmetry-adapted multi-determinantal wavefunctions}
    \label{sec:symmetry}
    \subsubsection{Symmetry groups, orbits, and representations}
      Consider a molecular system comprising $N_{\mathrm{e}}$ electrons placed in a fixed framework of $N_{\mathrm{n}}$ nuclei.
      In atomic units, the corresponding electronic Hamiltonian is given by
      \begin{equation}
        \hat{H} =
          \sum_{i=1}^{N_{\mathrm{e}}}
          -\frac{1}{2}\nabla_i^2
          + \sum_{i=1}^{N_{\mathrm{e}}} \sum_{j>i}^{N_{\mathrm{e}}}
          \frac{1}{\lvert \mathbf{r}_i - \mathbf{r}_j \rvert}
          + \sum_{i=1}^{N_{\mathrm{e}}} \sum_{A=1}^{N_{\mathrm{n}}}
        \frac{-Z_A}{\lvert \mathbf{r}_i - \mathbf{R}_A \rvert},
        \label{eq:H}
      \end{equation}
      where $\mathbf{r}_i$ denotes the position vector of the $i$\textsuperscript{th} electron and $\mathbf{R}_A$ that of the $A$\textsuperscript{th} nucleus with nuclear charge $Z_A$.
      This Hamiltonian determines the \textit{spatial unitary symmetry group} $\mathcal{G}$ of the system:
      \begin{equation}
        \mathcal{G} = \{ u : \hat{u} \hat{H} \hat{u}^{-1} = \hat{H} \},
      \end{equation}
      which consists of all \textit{spatial unitary symmetry operations} $\hat{u}$ that commute with $\hat{H}$.\cite{article:Irons2022,article:Huynh2024,article:Wibowo-Teale2024}
      Here, $u \in \mathcal{G}$ denotes an abstract group element, whereas $\hat{u}$ denotes its spatial unitary action on the $N_{\mathrm{e}}$-electron Hilbert space $\mathscr{H}$, on which $\hat{H}$ also acts. 

      Let us now consider a general state $\ket{\phi}$ on $\mathscr{H}$.
      The \textit{orbit}\cite{book:Smith2002} of this state generated by $\mathcal{G}$ is given by
      \begin{equation}
        \mathcal{G} \cdot \ket{\phi} = \{ \hat{u}_i \ket{\phi} : i = 1, \ldots, |\mathcal{G}|,\ u_i \in \mathcal{G} \},
      \end{equation}
      which comprises all symmetry-equivalent partners of $\ket{\phi}$ under the actions of $\mathcal{G}$ on $\mathscr{H}$.
      This orbit spans a certain subspace $\mathscr{F} \subseteq \mathscr{H}$; the decomposition of $\mathscr{F}$ in terms of the irreducible representations $\Gamma_i$ of $\mathcal{G}$ provides the \textit{spatial unitary symmetry} of $\ket{\phi}$:
      \begin{equation}
        \mathscr{F} = \bigoplus_{i} \Gamma_i^{\otimes k_i},
        \quad \dim \mathscr{F} \le |\mathcal{G}|,
        \label{eq:F-decomposition}
      \end{equation}
      where $k_i$ are the corresponding multiplicities.\cite{article:Huynh2024,article:Wibowo-Teale2024}
      The state $\ket{\phi}$ is said to be \textit{symmetry-conserved} with respect to $\mathcal{G}$ if $\mathscr{F}$ is irreducible (\textit{i.e.}, it contains only a single irreducible representation of $\mathcal{G}$ exactly once), and \textit{symmetry-broken} otherwise.\cite{article:Huynh2020}

    \subsubsection{Symmetry-orbit-based non-orthogonal configuration interaction (NOCI)}
      If $\ket{\phi}$ is symmetry-broken, it is possible to find linear combinations of the elements $\hat{u}_i \ket{\phi}$ of the orbit $\mathcal{G} \cdot \ket{\phi}$ that are symmetry-conserved.
      This can be achieved by allowing the $\hat{u}_i \ket{\phi}$ states to interact via the totally symmetric Hamiltonian $\hat{H}$, effectively solving a configuration interaction problem in the basis of $\mathcal{G} \cdot \ket{\phi}$.
      However, since the $\hat{u}_i \ket{\phi}$ states are in general non-orthogonal, this is a NOCI problem.\cite{article:Thom2009b,article:Sundstrom2014b,article:Huynh2020}
      The optimal linear combinations of the $\hat{u}_i \ket{\phi}$ states are obtained by solving the secular equation
      \begin{equation}
        \mathbf{H} \mathbf{C} = \mathbf{S} \mathbf{C} \mathbf{E},
      \end{equation}
      where $\mathbf{H}$ and $\mathbf{S}$ are the Hamiltonian and overlap matrices in the basis of $\mathcal{G} \cdot \ket{\phi}$ whose elements are given by
      \begin{equation}
        H_{ij} = \braket{\hat{u}_i \phi | \hat{H} | \hat{u}_j \phi},
        \qquad
        S_{ij} = \braket{\hat{u}_i \phi | \hat{u}_j \phi},
        \label{eq:noci-matelems}
      \end{equation}
      the columns of $\mathbf{C}$ contain the coefficients for the different optimal linear combinations of the $\hat{u}_i \ket{\phi}$ states, and the diagonal values of the diagonal matrix $\mathbf{E}$ give the corresponding optimal energies.
      Let us denote the $M$\textsuperscript{th} such optimal linear combination as $\ket{\Phi_M}$, refer to it as the \textit{$M$\textsuperscript{th} NOCI state}, and write
      \begin{equation}
        \ket{\Phi_M} = \sum_{i = 1}^{|\mathcal{G}|} \hat{u}_i \ket{\phi} C_{iM},
        \quad 1 \le M \le \dim \mathscr{F}.
        \label{eq:Phi-M-noci}
      \end{equation}
      Each multi-determinantal NOCI wavefunction $\ket{\Phi_M}$ is guaranteed to be symmetry-conserved in $\mathcal{G}$.

      Group closure can be exploited to reduce the number of matrix elements that have to be explicitly evaluated in Equation~\ref{eq:noci-matelems}.\cite{article:Huynh2024}
      Specifically, one finds
      \begin{alignat}{3}
        H_{ij}
          &= \braket{\hat{u}_i \phi | \hat{H} | \hat{u}_j \phi} \nonumber\\
          &= \braket{\hat{u}_i \phi | \hat{u}_j \hat{H} | \phi} && \qquad \textrm{$\hat{u}_j$ commutes with $\hat{H}$} \nonumber\\
          &= \braket{\hat{u}_j^{-1} \hat{u}_i \phi | \hat{H} | \phi} && \qquad \textrm{$\hat{u}_j$ is unitary} \nonumber\\
          &= \braket{\hat{u}_k \phi | \hat{H} | \phi} && \qquad u_k = u_j^{-1}u_i \nonumber\\
          &= H_{k1},
      \end{alignat}
      and, by an analogous argument,
      \begin{equation}
        S_{ij} = S_{k1},
      \end{equation}
      where $u_{1}$ is the identity in $\mathcal{G}$.
      Consequently, the computations of the matrices $\mathbf{H}$ and $\mathbf{S}$ scale as $O(|\mathcal{G}|)$, rather than $O(|\mathcal{G}|^2)$.
    
    \subsubsection{Symmetry projection}
      Consider the special case in which a given irreducible representation $\Gamma_i$ of $\mathcal{G}$ appears in $\mathscr{F}$ exactly once, \textit{i.e.}, $k_i = 1$ in Equation~\ref{eq:F-decomposition}.
      With this restriction, a symmetry-conserved multi-determinantal wavefunction with symmetry $\Gamma_i$ can be constructed efficiently by the use of the corresponding symmetry projection operator $\hat{P}_{\Gamma_i}$:\cite{book:Aktins2011}
      \begin{equation}
          \hat{P}_{\Gamma_i}
          = \frac{d_{\Gamma_i}}{|\mathcal{G}|} \sum_{j=1}^{|\mathcal{G}|} \chi_{\Gamma_i}(u_j)^* \hat{u}_j,
        \label{eq:proj-op}
      \end{equation}
      such that
      \begin{equation}
        \ket{\Phi_{\Gamma_i}}
          = \hat{P}_{\Gamma_i} \ket{\phi}
          = \frac{d_{\Gamma_i}}{|\mathcal{G}|} \sum_{j=1}^{|\mathcal{G}|} \chi_{\Gamma_i}(u_j)^* \hat{u}_j \ket{\phi},
        \label{eq:Phi-Gamma_i-proj}
      \end{equation}
      where $d_{\Gamma_i}$ is the dimension of $\Gamma_i$ and $\chi_{\Gamma_i}(u_j)$ the character of the group element $u_j$ in this irreducible representation.
      The projected wavefunction $\ket{\Phi_{\Gamma_i}}$ and the corresponding NOCI wavefunctions (Equation~\ref{eq:Phi-M-noci}) with the same symmetry belong to the same irreducible representation space; consequently, the former is a particular linear combination of the latter.
      Thus, in this case, NOCI and symmetry projection yield equivalent symmetry-conserved multi-determinantal wavefunctions.

      If, however, $\Gamma_i$ occurs in $\mathscr{F}$ more than once ($k_i > 1$), then the projected wavefunction $\ket{\Phi_{\Gamma_i}}$ remains symmetry-broken and the orbit $\mathcal{G} \cdot \ket{\Phi_{\Gamma_i}}$ spans a reducible space isomorphic to $\Gamma_i^{\otimes k_i}$.
      This arises because the projection operator $\hat{P}_{\Gamma_i}$ cannot distinguish between different, but isomorphic, copies of the same irreducible representation.
      In such cases, NOCI is required to construct fully symmetry-conserved multi-determinantal wavefunctions with symmetry $\Gamma_i$.

  \subsection{Generalised normal-ordered coupled-cluster (GNOCC) theory}
    \label{sec:gnocc}
    In this Section, we provide a brief review of GNOCC theory for the accurate determination of the ground and excited energies of $\hat{H}$; a complete account of the formalism can be found elsewhere.\cite{leeSpinfreeGeneralizedNormal2026}

    \subsubsection{Working equations}
      An operator product $\hat{A} \hat{B} \hat{C} \cdots$, where $\hat{A}, \hat{B}, \hat{C}, \ldots$  represent either creation or annihilation operators, is said to be in \emph{generalised normal order (GNO)}\cite{kutzelniggNormalOrderExtended1997,kutzelniggGeneralizedNormalOrdering2007} with respect to a reference wavefunction $\ket{\Phi}$ if
        \begin{equation}
          \braket{ \Phi | \{ \hat{A} \hat{B} \hat{C} \cdots \} | \Phi } = 0.
        \end{equation}
      We used braces $\{ \cdots \} $ to denote that the operator product has been placed in GNO.
      In GNOCC, the \textit{ansatz} is given by
      \begin{equation}
        \ket{\Psi} = \{ e^{\hat{T}} \} \ket{\Phi}
        \label{eqn:GNOCCwfn}
      \end{equation}
      where $\ket{\Phi}$ is an arbitrary \textit{spin} eigenfunction, \textit{i.e.}, $\hat{S}^2 \ket{\Phi} = S(S+1) \ket{\Phi}$ for $S = 0, \tfrac{1}{2}, 1, \tfrac{3}{2}, \ldots$, and $\hat{T}$ are excitation operators such that
      \begin{equation}
        \hat{T} = \sum_{pq} t^{q}_{p} \hat{E}^{p}_{q} + \frac{1}{2}\sum_{pqrs} t^{rs}_{pq}\hat{E}^{pq}_{rs} + \cdots = \sum_{\mu} t_{\mu} \hat{\tau}_{\mu}
      \end{equation}
      with $\hat{E}^{pq ... }_{rs ...}$ being spin-adapted excitation operators given by
      \begin{equation}
        \hat{E}^{pq \cdots}_{rs \cdots} =
        \sum_{\sigma, \tau, \ldots \in \{ \alpha, \beta\} }
          \hat{a}^{\dagger}_{p_{\sigma}} \hat{a}^{\dagger}_{q_{\tau}} \cdots
          \hat{a}_{s_{\tau}} \hat{a}_{r_{\sigma}},
      \end{equation}
      and $t^{pq \cdots}_{rs \cdots}$ being cluster amplitudes, which shall be denoted generically as $t_{\mu}$.
      We use Roman letters $p,q,r,s$ to denote \emph{spin-free} orbitals and Greek letters $\sigma, \tau$ to denote spin functions.
      This \textit{ansatz} ensures that $\ket{\Psi}$ is also a spin eigenfunction with the same $S$ quantum number as $\ket{\Phi}$.
      In this work, we adopt the singles-and-doubles approximation (GNOCCSD) where we retain only the single and double excitation operators such that
      \begin{equation}
        \hat{T} = \sum_{pq} t^{q}_{p} \hat{E}^{p}_{q} + \frac{1}{2}\sum_{pqrs} t^{rs}_{pq}\hat{E}^{pq}_{rs}.
      \end{equation}
    
      In the single-determinantal limit, the GNO exponential operator is equivalent to the standard exponential operator used in single-reference coupled cluster.
      For a multi-determinantal reference, the cluster operators $\hat{T}$ do not, in general, commute with each other.
      By using the GNO \textit{ansatz}, all $\hat{T}$ are placed within the same normal order and hence do not contract with each other by virtue of Wick's theorem.
      This greatly simplifies the working equation and allows for a finite truncation of the exponential operator. 
      
      In GNOCCSD, the correlation energy has the form
      \begin{equation}
        E_{\mathrm{corr}} = \bra{\Phi} \hat{H} \{ e^{\hat{T}} \} \ket{\Phi}.
        \label{eqn:Energy}
      \end{equation}
      The cluster amplitudes $t_{\mu}$ can be found by solving the residual equations
      \begin{equation}
        R_{\mu} = \bra{\Phi} \hat{\tau}_{\mu}^{\dagger} \hat{H} \{ e^{\hat{T}} \} \ket{\Phi}_{\mathrm{c}} = 0.
      \label{eqn:Residual}
      \end{equation}
      In the above equation, the subscript `c' indicates that only fully connected\cite{lyakhAlgebraicConnectivityAnalysis2012,lyakhAlgebraicConnectivityAnalysis2014} terms in the bra-ket $\bra{\Phi} \hat{\tau}_{\mu}^{\dagger} \hat{H} \{ e^{\hat{T}} \} \ket{\Phi}$ are considered.
      The residual equations are solved iteratively and the procedure is detailed in a previous publication.\cite{leeSpinfreeGeneralizedNormal2026}
  
      The contraction between operators in GNO naturally gives rise to one-body reduced density matrices ($1$-RDMs) and $k$-body cumulants.\cite{kutzelniggCumulantExpansionReduced1999,kutzelniggSpinfreeFormulationReduced2010,shamasundarCumulantDecompositionReduced2009}
      These $k$-body cumulants are, in turn, expressible in terms of $n$-body reduced density matrices ($n$-RDMs), where $n \leq k$.
      The reference state $\ket{\Phi}$ therefore comes into the energy and residual equations through its $n$-RDMs. 
      The construction of the required $n$-RDMs for this work will be detailed in Section \ref{sssection:NOCI2CAS}.

    \subsubsection{Redundancy handling procedure}
      \label{sec:redundancy}
      Linear dependencies amongst the residual equations plague MRCC methods because multiple excitations can lead to the same excited state onto which the Schr\"{o}dinger equation is projected.\cite{kohnStatespecificMultireferenceCoupledcluster2013}
      In the original formulation of GNOCCSD, the problem of linear dependency (redundancy) was resolved through the use of an orbital-dependent scheme.
      With this approach, GNOCCSD could be made to be numerically size-consistent at the expense of the sensitivity to the choice of molecular orbitals.
      However, an orbital-dependent scheme poses significant challenges when dealing with wavefunctions that exhibit high degrees of spatial degeneracy.
      This is because each of the spatially-degenerate wavefunctions can be expressed with a different molecular orbital basis, and an orbital-dependent correlation method will invariably find different correlation energies for different degenerate wavefunctions, thus artificially lifting their degeneracy.
      In this work, we circumvent this issue by resolving the linear dependency problem using the orbital-invariant canonical orthogonalisation method, which has been previously used in internally-contracted (ic) methods such as ic-MRCI\cite{menezesLocalCompleteActive2016} and icMRCC.\cite{evangelistaOrbitalinvariantInternallyContracted2011,hanauerPilotApplicationsInternally2011,blackEfficientImplementationInternally2023}
  
    In the canonical orthogonalisation approach, we first construct the \textit{excitation overlap matrix} $\mathbf{S}$ of dimensions $N_{\mathrm{exc}} \times N_{\mathrm{exc}}$ whose elements are
    \begin{equation}
      S_{\mu \nu} = \braket{ \Phi | \hat{\tau}_{\mu}^{\dagger} \hat{\tau}_{\nu} | \Phi },
    \end{equation}
    where $\hat{\tau}_{\mu}$ and $\hat{\tau}_{\nu}$ are cluster excitation operators.
    The excitation overlap matrix is canonically orthogonalised such that
    \begin{equation}
      (\mathbf{X}^{\dagger} \mathbf{S} \mathbf{X})_{ij} = \sum_{\mu \nu}  X^{*}_{i\mu} S_{\mu \nu} X_{\nu j} = \delta_{ij},
      \label{eqn:CanonicalOrth}
    \end{equation}
    where $\mathbf{X}$ is a rectangular matrix of dimensions $N_{\mathrm{exc}} \times N_{\mathrm{indep}}$, where $N_{\mathrm{indep}}$ is the number of linearly-independent excitations, corresponding to the number of eigenvectors of $\mathbf{S}$ with eigenvalues greater than a given \emph{redundancy threshold} $\epsilon$.
     The linearly-independent excitation basis used, $\tilde{\tau}_{i}$, is therefore given by
    \begin{equation}
      \tilde{\tau}_{i} = \sum_{\mu} \hat{\tau}_{\mu} X_{\mu i}.
      \label{eqn:Redundancy}
    \end{equation}

  \subsection{Spatial-symmetry-adapted multi-determinantal references for GNOCCSD}
    \subsubsection{Construction of complete-active-space (CAS)-type wavefunctions}
      \label{sssection:NOCI2CAS}
      The GNOCCSD method assumes a complete-active-space (CAS)-type wavefunction, in which all constituent determinants share a common set of molecular orbitals but differ in their occupation patterns.
      The orbitals characterising its constituent determinants can be partitioned into three classes: \textit{core (c)} (doubly occupied in all determinants), \textit{virtual (v)} (unoccupied in all determinants), and \textit{active (a)} (variable occupation numbers on the interval $(0, 2)$).
      We note that \textit{core} orbitals as defined in this work can be equivalently labelled \textit{inactive}.
      Such a partitioning is desirable because it allows us to express the GNOCCSD working equations as a function of $n$-RDMs in the basis of active orbitals only.
      In contrast, a general multi-determinantal wavefunction is usually expressed in terms of determinants built from different molecular orbital bases.
      For the purpose of this work, it is desirable to transform such wavefunctions into a common orbital basis.
      A natural choice is the natural orbital basis, obtained by diagonalising a suitably chosen $1$-RDM corresponding to the multi-determinantal wavefunction, whose eigenvalues give occupation numbers that lie in the interval $[0, 2]$.
      
      Since GNOCCSD exhibits coupled-cluster singles and doubles (CCSD) scaling [$\mathcal{O}(N^6)$] when the number of active orbitals is small relative to the number of core and virtual orbitals, it is advantageous to limit the size of the active space.
      In this work, we consider any natural orbitals with an occupation number larger than 1.99 as \textit{core (c)}, and those with an occupation number smaller than $0.01$ as \textit{virtual (v)}.
      All other natural orbitals are \textit{active (a)}.

      \begin{figure}[h!]
        \centering
        \includegraphics[width=0.8\linewidth]{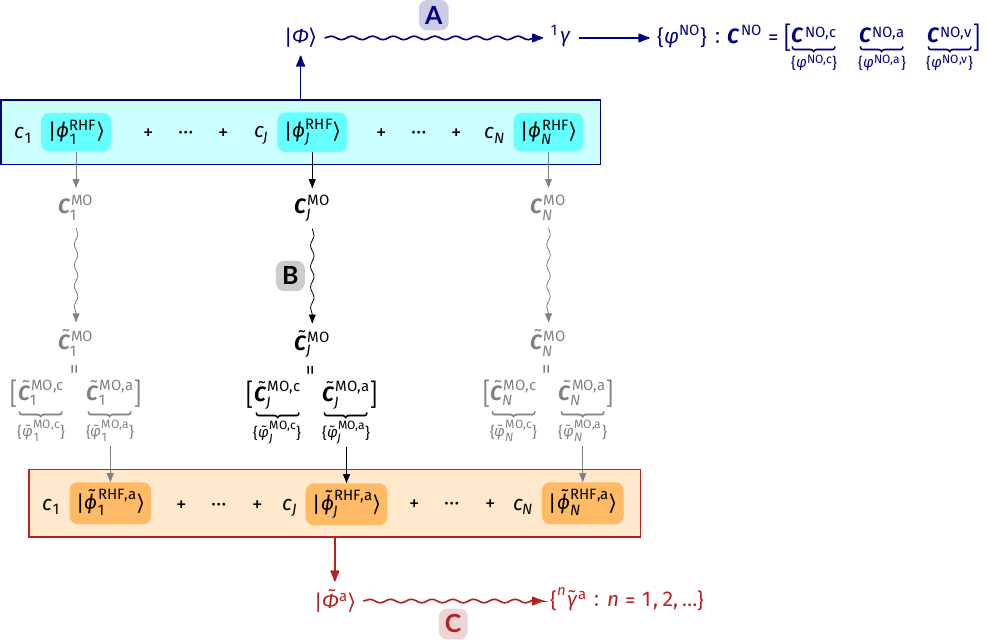}
        \caption{%
          Construction of an active CAS-type wavefunction $\ket{\tilde{\Phi}^{\mathrm{a}}}$ (orange box) from a general multi-determinantal wavefunction $\ket{\Phi}$ (cyan box).
          \textbf{A}. A suitable $1$-RDM $\prescript{1}{}{\gamma}$ for $\ket{\Phi}$ is constructed and diagonalised to obtain a set of natural orbitals $\{\varphi^{\mathrm{NO}}\}$ with coefficient matrix $\mathbf{C}^{\mathrm{NO}}$ which can be partitioned into the \textit{core} part $\mathbf{C}^{\mathrm{NO,c}} \Leftrightarrow \{\varphi^{\mathrm{NO,c}}\}$, the \textit{active} part $\mathbf{C}^{\mathrm{NO,a}} \Leftrightarrow \{\varphi^{\mathrm{NO,a}}\}$, and the \textit{virtual} part $\mathbf{C}^{\mathrm{NO,v}} \Leftrightarrow \{\varphi^{\mathrm{NO,v}}\}$.
          \textbf{B}. For each constituent RHF determinant $\ket{\phi_J^{\mathrm{RHF}}}$ in $\ket{\Phi}$, the coefficient matrix $\mathbf{C}_J^{\mathrm{MO}}$ is, after an SVD procedure, unitarily transformed into $\tilde{\mathbf{C}}_J^{\mathrm{MO}}$ which contains a \textit{core} part $\tilde{\mathbf{C}}_J^{\mathrm{MO,c}} \Leftrightarrow \{\tilde{\varphi}_J^{\mathrm{MO,c}}\}$ whose span closely resembles that of $\{\varphi^{\mathrm{NO,c}}\}$, and an \textit{active} part $\tilde{\mathbf{C}}_J^{\mathrm{MO,a}} \Leftrightarrow \{\tilde{\varphi}_J^{\mathrm{MO,a}}\}$ that forms the corresponding active RHF determinant $\ket{\tilde{\phi}_i^{\mathrm{RHF,a}}}$.
          \textbf{C}. The active RHF determinants in turn constitute the active CAS-type multi-determinantal wavefunction $\ket{\tilde{\Phi}^{\mathrm{a}}}$, which provides a description of the electronic structure within the active space via the active $n$-RDMs $\prescript{n}{}{\tilde{\gamma}}^{\mathrm{a}}$.
        }
        \label{fig:cas-type-construction}
      \end{figure}
      
      Figure~\ref{fig:cas-type-construction} illustrates the workflow for the construction of a CAS-type wavefunction from a general multi-determinantal wavefunction.
      This workflow comprises three main steps.

      \paragraph{Step A. Determination of the natural orbitals of the multi-determinantal wavefunction.}
        With the above classification of natural orbitals based on their occupation numbers, the constituent determinants of a general multi-determinantal wavefunction $\ket{\Phi}$ may be approximately expressed in a common natural orbital basis.
        To this end, let $\ket{\Phi}$ be constructed from $N$ Restricted Hartree--Fock (RHF) wavefunctions $\ket{\phi^{\mathrm{RHF}}_1}, \ldots, \ket{\phi^{\mathrm{RHF}}_J}, \ldots, \ket{\phi^{\mathrm{RHF}}_N}$, where each $\ket{\phi^{\mathrm{RHF}}_J}$ comprises $N_{\mathrm{occ}}$ doubly occupied RHF molecular orbitals $\{\varphi^{\mathrm{MO}}_J\}$ with coefficient matrix $\mathbf{C}^{\mathrm{MO}}_J$.
        An appropriate $1$-RDM $\prescript{1}{}{\gamma}$ is first constructed for $\ket{\Phi}$ (\textit{cf.} Section~\ref{sec:sym-RDMs} for details regarding the required symmetry considerations for $\prescript{1}{}{\gamma}$), and then diagonalised to give a set of natural orbitals $\{\varphi^{\mathrm{NO}}\}$ with coefficient matrix $\mathbf{C}^{\mathrm{NO}}$.
        The resulting natural orbitals are then partitioned into three sets according to their occupation numbers as \textit{core} $\mathbf{C}^{\mathrm{NO,c}} \Leftrightarrow \{\varphi^{\mathrm{NO,c}}\}$, \textit{active} $\mathbf{C}^{\mathrm{NO,a}} \Leftrightarrow \{\varphi^{\mathrm{NO,a}}\}$, and \textit{virtual} $\mathbf{C}^{\mathrm{NO,v}} \Leftrightarrow \{\varphi^{\mathrm{NO,v}}\}$.
        Denoting the number of core natural orbitals by $N_{\mathrm{core}}$, we can easily see that $N_{\mathrm{occ}} \ge N_{\mathrm{core}}$.

      \paragraph{Step B. Identification of the common core space.}
        For each $\ket{\phi^{\mathrm{RHF}}_J}$, we seek a unitary transformation of its occupied RHF molecular orbitals $\{\varphi^{\mathrm{MO}}_J\}$ such that $N_{\mathrm{core}}$ of the transformed orbitals best approximate the core natural orbitals $\{\varphi^{\mathrm{NO,c}}\}$.
        This can be achieved via a singular value decomposition (SVD) of the $N_{\mathrm{occ}} \times N_{\mathrm{core}}$ overlap matrix between the occupied RHF molecular orbitals and the core natural orbitals, 
        \begin{equation}
          \mathbf{S}^{\mathrm{MO,NO}}_J
            = \mathbf{C}^{\mathrm{MO} \dagger}_J \mathbf{S}^{\mathrm{AO}} \mathbf{C}^{\mathrm{NO,c}}
            = \mathbf{U}_J \boldsymbol{\Sigma}_J \mathbf{V}^{\dagger}_J,
        \end{equation}
        where $\mathbf{S}^{\mathrm{AO}}$ is the atomic orbital overlap matrix, $\mathbf{U}_J$ is an $N_{\mathrm{occ}} \times N_{\mathrm{occ}}$ unitary matrix, $\mathbf{V}_J$ is an $N_{\mathrm{core}} \times N_{\mathrm{core}}$ unitary matrix, and $\boldsymbol{\Sigma}_J$ is a rectangular diagonal matrix of dimension $N_{\mathrm{occ}} \times N_{\mathrm{core}}$.
        The transformed occupied molecular orbital coefficient matrix $\tilde{\mathbf{C}}^{\mathrm{MO}}_J$, which contains $N_{\mathrm{occ}}$ columns, is then given by
        \begin{equation}
          \tilde{\mathbf{C}}^{\mathrm{MO}}_J = \mathbf{C}^{\mathrm{MO}}_J \mathbf{U}_J,
        \end{equation}
        which defines the transformed RHF determinant $\ket{\tilde{\phi}^{\mathrm{RHF}}_J}$. 
        Owing to the invariance of a Slater determinant under unitary transformations within the occupied subspace, $\ket{\phi^{\mathrm{RHF}}_J}$ and $\ket{\tilde{\phi}^{\mathrm{RHF}}_J}$ are identical.
        Furthermore, the $N_{\mathrm{core}}$ columns of $\tilde{\mathbf{C}}^{\mathrm{MO}}_J$ that correspond to the $N_{\mathrm{core}}$ largest singular values give the best approximation to the core natural orbitals that can be constructed from the occupied molecular orbitals of $\ket{\phi^{\mathrm{RHF}}_J}$.
        These columns are collected into the rectangular matrix $\tilde{\mathbf{C}}^{\mathrm{MO},\mathrm{c}}_J$, which define the set of \textit{occupied} RHF molecular orbitals $\{\tilde{\varphi}_J^{\mathrm{MO,c}}\}$ whose span closely resembles that of $\{\varphi^{\mathrm{NO,c}}\}$.
  
        In general, different $\ket{\phi^{\mathrm{RHF}}_J}$ wavefunctions in $\ket{\Phi}$ give rise to distinct approximations $\tilde{\mathbf{C}}^{\mathrm{MO,c}}_J \Leftrightarrow \{\tilde{\varphi}_J^{\mathrm{MO,c}}\}$ to the core natural orbitals.
        Consequently, the above procedure does not, in general, yield an exact common orbital basis for all constituent determinants of a multi-determinantal wavefunction.
        However, when the determinants $\ket{\phi^{\mathrm{RHF}}_J}$ form a complete symmetry orbit, which can be ensured by construction (\textit{cf.} Section~\ref{sec:symmetry}), the core natural orbitals $\{\varphi^{\mathrm{NO,c}}\}$ of $\ket{\Phi}$ are expected to be approximately invariant under the symmetry operations of the system. 
        We emphasise that this expectation is motivated based on chemical intuition rather than a mathematical argument.
        Equivalently, a determinant constructed from $\{\varphi^{\mathrm{NO,c}}\}$ would be approximately totally symmetric with respect to the symmetry group $\mathcal{G}$.
        Furthermore, in most chemically relevant cases, each constituent determinant $\ket{\phi^{\mathrm{RHF}}_J}$ is expected to contain a closed-shell core component whose span is similar to that of the core natural orbitals.
        It follows that the resulting $\tilde{\mathbf{C}}^{\mathrm{MO,c}}_J \Leftrightarrow \{\tilde{\varphi}_J^{\mathrm{MO,c}}\}$ from the SVD procedure above are able to describe, to a good approximation, nearly identical core spaces, which are themselves very similar to that defined by $\mathbf{C}^{\mathrm{NO,c}} \Leftrightarrow \{\varphi^{\mathrm{NO,c}}\}$.
        In other words,
        \begin{equation}
          \spn \{\tilde{\varphi}_1^{\mathrm{MO,c}}\}
          \simeq \cdots
          \simeq \spn \{\tilde{\varphi}_J^{\mathrm{MO,c}}\}
          \simeq \cdots
          \simeq \spn \{\tilde{\varphi}_N^{\mathrm{MO,c}}\}
          \simeq \spn \{\varphi^{\mathrm{NO,c}}\}.
          \label{eq:core-spans}
        \end{equation}

      \paragraph{Step C. Correlation in the active space.}
        The remaining $N_{\mathrm{occ}} - N_{\mathrm{core}}$ columns of $\tilde{\mathbf{C}}^{\mathrm{MO}}_J$, \textit{i.e.}, those not included in $\tilde{\mathbf{C}}^{\mathrm{MO},\mathrm{c}}_J$, are collected into $\tilde{\mathbf{C}}^{\mathrm{MO},\mathrm{a}}_J$, which defines the set $\{\tilde{\varphi}_J^{\mathrm{MO,a}}\}$ which we shall \textit{designate} as the \textit{active} RHF molecular orbitals.
        The space spanned by this set may be interpreted analogously as the best approximation, within the occupied molecular orbital space of $\ket{\phi^{\mathrm{RHF}}_J}$, to the full active space defined by $\mathbf{C}^{\mathrm{NO,a}} \Leftrightarrow \{\varphi^{\mathrm{NO,a}}\}$. 
        We note, however, that each individual active subspace described by $\tilde{\mathbf{C}}^{\mathrm{MO},\mathrm{a}}_J \Leftrightarrow \{\tilde{\varphi}_J^{\mathrm{MO,a}}\}$ must, in general, be contained within the full active space.
        This is partly because $N_{\mathrm{occ}} - N_{\mathrm{core}} \le N_{\mathrm{active}}$, where $N_{\mathrm{active}}$ denotes the number of active natural orbitals in $\{\varphi^{\mathrm{NO,a}}\}$, owing to the fractionality of active natural occupation numbers.
        The full active space can be recovered by taking the union of the subspaces spanned by the $\tilde{\mathbf{C}}^{\mathrm{MO},\mathrm{a}}_J \Leftrightarrow \{\tilde{\varphi}_J^{\mathrm{MO,a}}\}$, \textit{i.e.},
        \begin{equation}
          \spn \{\varphi^{\mathrm{NO,a}}\} \simeq \bigcup_{J=1}^{N} \spn \{\tilde{\varphi}_J^{\mathrm{MO,a}}\}.
          \label{eq:active-spans}
        \end{equation}

        We now seek to determine the $n$-RDMs describing the electrons in the active space, expressed in the basis of the active natural orbitals, $\{\varphi^{\mathrm{NO,a}}\}$. 
        To this end, we define an \textit{active} CAS-type multi-determinantal wavefunction $\ket{\tilde{\Phi}^{\mathrm{a}}}$ as a linear combination of the corresponding \textit{active} RHF determinants $\ket{\tilde{\phi}^{\mathrm{RHF,a}}_J}$, each defined by $\tilde{\mathbf{C}}^{\mathrm{MO},\mathrm{a}}_J \Leftrightarrow \{\tilde{\varphi}_J^{\mathrm{MO,a}}\}$ and weighted by the coefficients $c_J$ of the original multi-determinantal wavefunction $\ket{\Phi}$.
        The wavefunction $\ket{\tilde{\Phi}^{\mathrm{a}}}$ thus provides a description of the electronic structure within the active space, from which the appropriate \textit{active} $n$-RDMs $\prescript{n}{}{\tilde{\gamma}}^{\mathrm{a}}$ may be constructed in the basis of $\{\varphi^{\mathrm{NO,a}}\}$ (\textit{cf.} Section~\ref{sec:sym-RDMs} for details regarding the required symmetry considerations for the $n$-RDMs).
        The active $n$-RDMs are then utilised in GNOCCSD calculations as detailed in Section~\ref{sec:gnocc}.

      \paragraph{Exactness of the CAS-type construction.}
        The above procedure is only exact if equality is attained in Equations~\ref{eq:core-spans} and \ref{eq:active-spans}.
        A simple way to verify the exactness of this procedure is to construct the \textit{full} CAS-type multi-determinantal wavefunction $\ket{\tilde{\Phi}}$.
        For each constituent determinant $\ket{\phi_J^{\mathrm{RHF}}}$ in the original multi-determinantal wavefunction $\ket{\Phi}$, we define the corresponding CAS-type determinant $\ket{\tilde{\phi}^{\mathrm{CAS}}_J}$ whose coefficient matrix is given by
        \begin{equation}
          \begin{bmatrix}
            \smash[b]{\underbrace{\mathbf{C}^{\mathrm{NO,c}}_{\vphantom{J}}}_{\{ \varphi^{\mathrm{NO,c}} \}}}
            &
            \smash[b]{\underbrace{\tilde{\mathbf{C}}^{\mathrm{MO,a}}_{J}}_{\{ \tilde{\varphi}^{\mathrm{MO,a}}_J \}}}
          \end{bmatrix}.
          \vspace{0.5cm}
        \end{equation}
        Therefore, each $\ket{\tilde{\phi}^{\mathrm{CAS}}_J}$ is effectively $\ket{\tilde{\phi}^{\mathrm{RHF}}_J}$ but with the core natural orbital approximation $\tilde{\mathbf{C}}^{\mathrm{MO,c}}_J \Leftrightarrow \{\tilde{\varphi}_J^{\mathrm{MO,c}}\}$ replaced by the actual core natural orbitals $\mathbf{C}^{\mathrm{NO,c}} \Leftrightarrow \{\varphi^{\mathrm{NO,c}}\}$ themselves.
        The full $\ket{\tilde{\Phi}}$ is then a linear combination of $\ket{\tilde{\phi}^{\mathrm{CAS}}_J}$, each weighted by the coefficients $c_J$ of the original $\ket{\Phi}$.
        The exactness of $\ket{\tilde{\Phi}}$ compared to $\ket{\Phi}$ can then be quantified based on the energy difference
        \begin{equation}
          E_{\ket{\tilde{\Phi}}} - E_{\ket{\Phi}}.
          \label{eq:cas-error}
        \end{equation}

    \subsubsection{Symmetry considerations for $n$-RDMs of degenerate wavefunctions}
      \label{sec:sym-RDMs}

      A key aspect in the above procedure is the construction of $n$-RDMs for various multi-determinantal wavefunctions, together with the diagonalisation of $1$-RDMs to obtain natural orbitals that enable the identification of core and active spaces.
      If these wavefunctions are degenerate, the $n$-RDMs obtained in a na\"{i}ve manner generally break symmetry and consequently contain unphysical components that not only lead to incorrect descriptions of the electronic structure, but also introduce significant challenges for the convergence of GNOCCSD calculations.
      It is therefore vital to construct the $n$-RDMs in such a way that respects the symmetry of the system.
      In the following, we outline the considerations required to achieve this.
      
      \paragraph{Symmetry of pure-state $n$-RDMs.}
        Consider a set of $d$-fold degenerate wavefunctions
        \begin{equation}
          \mathcal{B} = \{
             \ket{\Phi_i}
             :
             i = 1, \ldots, d
          \}
          \label{eq:basis-d}
        \end{equation}
        that forms a basis for a $d$-dimensional irreducible representation $\Gamma$ of the spatial unitary symmetry group $\mathcal{G}$ of the system.
        The transformation of these wavefunctions under the action of the elements in $\mathcal{G}$ is given by
        \begin{equation}
          \hat{u} \ket{\Phi_i}
          = \sum_{i'=1}^{d}
            \ket{\Phi_{i'}}
            D_{i'i}^{\Gamma}(u),
          \label{eq:deg-repmat}
        \end{equation}
        where $\mathbf{D}^{\Gamma}(u)$ is the representation matrix of $u$ in $\Gamma$ in the basis $\mathcal{B}$.
        The \textit{pure-state} $n$-RDMs corresponding to the wavefunctions in $\mathcal{B}$ have the form\cite{book:Parr1989}
        \begin{equation}
          \prescript{n}{}{\gamma}_i(\mathbf{x}'_1 \mathbf{x}'_2 \cdots \mathbf{x}'_n, \mathbf{x}_1 \mathbf{x}_2 \cdots \mathbf{x}_n)
          = \begin{pmatrix} N_{\mathrm{e}} \\ n \end{pmatrix}
            \int \Phi_i(%
              \mathbf{x}'_1 \mathbf{x}'_2 \cdots \mathbf{x}'_n \mathbf{x}_{n+1} \cdots \mathbf{x}_{N_{\mathrm{e}}}
            ) \ \Phi_i^*(%
              \mathbf{x}_1 \mathbf{x}_2 \cdots \mathbf{x}_n \mathbf{x}_{n+1} \cdots \mathbf{x}_{N_{\mathrm{e}}}
            ) \ \D\mathbf{x}_{n+1} \cdots \D\mathbf{x}_{N_{\mathrm{e}}}.
        \end{equation}
        Let us also define the \textit{transition} $n$-RDMs between the wavefunctions in $\mathcal{B}$ as
        \begin{equation}
          \prescript{n}{}{\gamma}_{ij}(\mathbf{x}'_1 \mathbf{x}'_2 \cdots \mathbf{x}'_n, \mathbf{x}_1 \mathbf{x}_2 \cdots \mathbf{x}_n)
          = \begin{pmatrix} N_{\mathrm{e}} \\ n \end{pmatrix}
            \int \Phi_i(%
              \mathbf{x}'_1 \mathbf{x}'_2 \cdots \mathbf{x}'_n \mathbf{x}_{n+1} \cdots \mathbf{x}_{N_{\mathrm{e}}}
            ) \ \Phi_j^*(%
              \mathbf{x}_1 \mathbf{x}_2 \cdots \mathbf{x}_n \mathbf{x}_{n+1} \cdots \mathbf{x}_{N_{\mathrm{e}}}
            ) \ \D\mathbf{x}_{n+1} \cdots \D\mathbf{x}_{N_{\mathrm{e}}}.
        \end{equation}
        The symmetry transformations of the transition $n$-RDMs $\prescript{n}{}{\gamma}_{ij}$ can thus be written as
        \begin{align}
          \hat{u} \prescript{n}{}{\gamma}_{ij}(\mathbf{x}'_1 \mathbf{x}'_2 \cdots \mathbf{x}'_n, \mathbf{x}_1 \mathbf{x}_2 \cdots \mathbf{x}_n)
          &= \begin{pmatrix} N_{\mathrm{e}} \\ n \end{pmatrix}
            \int \hat{u} \Phi_i(%
              \mathbf{x}'_1 \mathbf{x}'_2 \cdots \mathbf{x}'_n \mathbf{x}_{n+1} \cdots \mathbf{x}_{N_{\mathrm{e}}}
            ) \ \hat{u} \Phi_j^*(%
              \mathbf{x}_1 \mathbf{x}_2 \cdots \mathbf{x}_n \mathbf{x}_{n+1} \cdots \mathbf{x}_{N_{\mathrm{e}}}
            ) \ \D\mathbf{x}_{n+1} \cdots \D\mathbf{x}_{N_{\mathrm{e}}} \nonumber\\
          &= \sum_{i'j'}^{d}
            D_{i'i}^{\Gamma}(u) D_{j'j}^{\Gamma}(u)^*
            \begin{pmatrix} N_{\mathrm{e}} \\ n \end{pmatrix}
            \int \Phi_{i'}(%
              \mathbf{x}'_1 \mathbf{x}'_2 \cdots \mathbf{x}'_n \mathbf{x}_{n+1} \cdots \mathbf{x}_{N_{\mathrm{e}}}
            ) \ \Phi_{j'}^*(%
              \mathbf{x}_1 \mathbf{x}_2 \cdots \mathbf{x}_n \mathbf{x}_{n+1} \cdots \mathbf{x}_{N_{\mathrm{e}}}
            ) \ \D\mathbf{x}_{n+1} \cdots \D\mathbf{x}_{N_{\mathrm{e}}} \nonumber\\
          &= \sum_{i'j'}^{d}
            \prescript{n}{}{\gamma}_{i'j'}(\mathbf{x}'_1 \mathbf{x}'_2 \cdots \mathbf{x}'_n, \mathbf{x}_1 \mathbf{x}_2 \cdots \mathbf{x}_n)
            D_{i'i}^{\Gamma}(u) D_{j'j}^{\Gamma}(u)^*.
        \end{align}
        It follows that the set of transition $n$-RDMs $\{\prescript{n}{}{\gamma}_{ij} : i, j = 1, \ldots, d\}$ forms a basis for the direct-product representation $\Gamma \otimes \Gamma^*$.
  
        In this work, we only consider real-valued wavefunctions describing even numbers of electrons, so that $\Gamma$ is necessarily a real and single-valued irreducible representation, and $\Gamma$ is thus identical to $\Gamma^*$.
        In addition, since $\prescript{n}{}{\gamma}_i = \prescript{n}{}{\gamma}_{ii}$ and is therefore trivially symmetric with respect to the exchange of the two occurrences of $\Phi_i$ in the integrand, the pure-state $n$-RDMs must be part of a basis spanning the symmetrised square of $\Gamma$, which we shall denote $\Gamma^2_{+}$.
        If $d = 1$ and $\Gamma$ is real and non-degenerate, then $\Gamma^2_{+}$ is simply the totally symmetric irreducible representation of $\mathcal{G}$; in this case, the pure-state $n$-RDMs are totally symmetric.
        However, if $d > 1$, then $\Gamma^2_{+}$ is necessarily reducible (and must also contain the totally symmetric irreducible representation); consequently, the pure-state $n$-RDMs are symmetry-broken.

      \paragraph{Totally symmetric ensemble $n$-RDMs.}
        The spin-traced diagonal elements of the pure-state $n$-RDMs,
        \begin{equation}
          \prescript{n}{}{\rho}(\mathbf{r}_1 \mathbf{r}_2 \cdots \mathbf{r}_n)
          = \int \prescript{n}{}{\gamma}_i(\mathbf{x}_1 \mathbf{x}_2 \cdots \mathbf{x}_n, \mathbf{x}_1 \mathbf{x}_2 \cdots \mathbf{x}_n) \ \D s_1 \D s_2 \cdots \D s_n
          \label{eq:spintraced-n-rdm}
        \end{equation}
        are well-established physical observables that are also scalar quantities [\textit{e.g.}, $\prescript{1}{}{\rho}(\mathbf{r})$ corresponds to the electron density and $\prescript{2}{}{\rho}(\mathbf{x}_1 \mathbf{x}_2)$ to the pair density].
        Consequently, they are required to be totally symmetric as they need to remain invariant under all symmetry transformations of the underlying system.
        However, as discussed above, the pure-state $n$-RDMs associated with degenerate wavefunctions are symmetry-broken.
        It follows that their spin-traced diagonals contain not only the physically meaningful totally symmetric component, but also spurious contributions that transform according to non-totally-symmetric representations.
        It is therefore desirable to retain only the totally symmetric component of the $n$-RDMs associated with degenerate wavefunctions.

        To this end, we employ the projection operator (Equation~\ref{eq:proj-op}) for the totally symmetric irreducible representation $\Gamma_{\mathrm{tot.sym.}}$ to extract the desired totally symmetric component of the $n$-RDMs:
        \begin{equation}
          \prescript{n}{}{\bar{\gamma}}(\mathbf{x}'_1 \mathbf{x}'_2 \cdots \mathbf{x}'_n, \mathbf{x}_1 \mathbf{x}_2 \cdots \mathbf{x}_n)
            = \hat{P}_{\Gamma_{\mathrm{tot.sym.}}} \prescript{n}{}{\gamma}_i(\mathbf{x}'_1 \mathbf{x}'_2 \cdots \mathbf{x}'_n, \mathbf{x}_1 \mathbf{x}_2 \cdots \mathbf{x}_n)
            = \frac{1}{|\mathcal{G}|}
                \sum_{j=1}^{|\mathcal{G}|} \hat{u}_j
                  \prescript{n}{}{\gamma}_i(\mathbf{x}'_1 \mathbf{x}'_2 \cdots \mathbf{x}'_n, \mathbf{x}_1 \mathbf{x}_2 \cdots \mathbf{x}_n).
          \label{eq:P-n_gamma_i}
        \end{equation}
        By construction, $\prescript{n}{}{\bar{\gamma}}$ is a convex combination over the orbit $\mathcal{G} \cdot \prescript{n}{}{\gamma}_i$ (\textit{i.e.}, over all symmetry-equivalent partners of $\prescript{n}{}{\gamma}_i$).
        It therefore defines an ensemble $n$-RDM that is strictly invariant under all symmetry operations of the system.\cite{book:Parr1989}
        
        Moreover, the Great Orthogonality Theorem\cite{book:Aktins2011} can be invoked to show that $\prescript{n}{}{\bar{\gamma}}$ can be alternatively constructed as an equal-weight ensemble of pure-state $n$-RDMs associated with the degenerate wavefunctions in the basis $\mathcal{B}$ (Equation~\ref{eq:basis-d}):
        \begin{equation}
          \prescript{n}{}{\bar{\gamma}} = \frac{1}{d} \sum_{i=1}^{d} \prescript{n}{}{\gamma}_i,
          \label{eq:equi-ensemble}
        \end{equation}
        which also demonstrates that $\prescript{n}{}{\bar{\gamma}}$ is independent of the particular pure-state $n$-RDM to which the projection operator is applied in Equation~\ref{eq:P-n_gamma_i}.
        It can further be shown, by making use of the Little Orthogonality Theorem,\cite{book:Aktins2011} that
        \begin{equation}
          \int
            \hat{P}_{\Gamma}
            \prescript{n}{}{\gamma}_i(\mathbf{x}_1 \mathbf{x}_2 \cdots \mathbf{x}_n, \mathbf{x}_1 \mathbf{x}_2 \cdots \mathbf{x}_n) \ \D\mathbf{x}_1 \D\mathbf{x}_2 \cdots \mathbf{x}_n
          = \begin{cases}
            \begin{pmatrix}
              N_{\mathrm{e}} \\ n
            \end{pmatrix} & \mathrm{if} \ \Gamma \sim \Gamma_{\mathrm{tot.sym.}},
            \\[18pt]
            0 & \mathrm{otherwise},
          \end{cases}
        \end{equation}
        indicating that only the totally symmetric component of the $n$-RDM contains the correct number of $n$-tuple electrons, whereas non-totally-symmetric components carry no electrons at all and are thus unphysical.

        In light of these considerations, the $n$-RDMs constructed in steps \textbf{A} and \textbf{C} of Figure~\ref{fig:cas-type-construction} must therefore be suitably symmetrised.
        This can be achieved either by the use of the totally symmetric projection operator as in Equation~\ref{eq:P-n_gamma_i}, or by constructing the equal-weight ensemble as in Equation~\ref{eq:equi-ensemble}.

    \subsubsection{Residual spatial symmetry breaking}
      \label{sec:residualspatialsymbreaking}
      The aforementioned symmetry adaptation does not preclude the overall GNOCCSD wavefunction from exhibiting residual symmetry breaking, however small. 
      Consider the GNOCCSD wavefunction \textit{ansatz} (Equation~\ref{eqn:GNOCCwfn}) where the reference state $\ket{\Phi}$ has been symmetry-adapted such that it transforms as some irreducible representation $\Gamma$ of the unitary symmetry group $\mathcal{G}$.
      For the GNOCCSD wavefunction $\ket{\Psi}$ to be symmetry-conserved and also transform as $\Gamma$, the GNO exponential operator $\{ e^{\hat{T}} \}$ must be totally symmetric with respect to $\mathcal{G}$.
      Since $\{ e^{\hat{T}} \} = \{ \sum_{n=0}^{\infty} \frac{1}{n!}\hat{T}^{n}\}$, $\{ e^{\hat{T}} \}$ transforms as the totally symmetric irreducible representation when $\hat{T}$ does too.
      We shall show that this is not the case for $\hat{T}$ in general.
      For $\hat{T}$ to be totally symmetric, we must have
      \begin{equation}
          \hat{T} = \hat{u}\hat{T}\hat{u}^{-1}  \quad \forall \hat{u} \in \mathcal{G}.
      \end{equation}
      However, if we take, for example, $\hat{T} = \sum_{i\in C} \sum_{a \in V} t^{i}_{a} \hat{a}^{\dagger}_{a} \hat{a}_{i}$, where $C$ and $V$ denote the sets of core and virtual indices, respectively, we find that
      \begin{equation}
        \begin{split}
          \hat{u}\hat{T}\hat{u}^{-1}
          &= \hat{u} \left(\sum_{i\in C} \sum_{a \in V} t^{i}_{a} \hat{a}^{\dagger}_{a} \hat{a}_{i}\right) \hat{u}^{-1} \\
          &= \sum_{i\in C} \sum_{a \in V} t^{i}_{a}
            \ \hat{u} \hat{a}^{\dagger}_{a} \hat{u}^{-1}
            \ \hat{u} \hat{a}_{i} \hat{u}^{-1} \\
          &= \sum_{i\in C} \sum_{a \in V} t^{i}_{a}
            \ \sum_{p} \hat{a}^{\dagger}_{p} D^{*}_{pa}(u)
            \ \sum_{q} \hat{a}_{q} D_{qi}(u) \\
          &= \sum_{pq} \hat{a}^{\dagger}_{p} \hat{a}_{q} \sum_{i\in C} \sum_{a \in V} t^{i}_{a} D^{*}_{pa}(u)  D_{qi}(u) \\
          &= \sum_{pq} t^{q}_{p}(u) \hat{a}^{\dagger}_{p} \hat{a}_{q},
        \end{split}
      \end{equation}
      where $\mathbf{D}(u)$ is the representation matrix of the element $u \in \mathcal{G}$ in the \textit{full} (core, active, and virtual) orbital basis.
      In the last step, we identified $t^{q}_{p}(u) = \sum_{i\in C} \sum_{a \in V} t^{i}_{a} D^{*}_{pa}(\hat{u}) D_{qi}(\hat{u})$.
      Therefore, we require 
      \begin{equation}
        \sum_{i\in C} \sum_{a \in V} t^{i}_{a} \hat{a}^{\dagger}_{a} \hat{a}_{i} = \sum_{pq} t^{q}_{p} \hat{a}^{\dagger}_{p} \hat{a}_{q}
      \end{equation}
      for $\hat{T}$ to be totally symmetric.
      However, since $p$ and $q$ run over all orbitals in the full basis, this equality does not hold in general.
      This example thus illustrates that the GNOCCSD wavefunction can still be symmetry-broken in $\mathcal{G}$ due to the symmetry breaking of $\{ e^{\hat{T}} \}$, even if the reference state $\ket{\Phi}$ is symmetry-conserved. 
      Nevertheless, the proposed method still provides a way for targeting a specific excited state that transforms as a particular irreducible representation, thereby enabling symmetry-resolved characterisation of a correlated electronic state.
      
    \section{Computational Details}
      \label{sec:compdetails}
      \subsection{Molecular systems}
\begin{figure}[h!]
            \centering
            \includegraphics[width=0.55\linewidth, trim=0 9cm 0 9cm, clip]{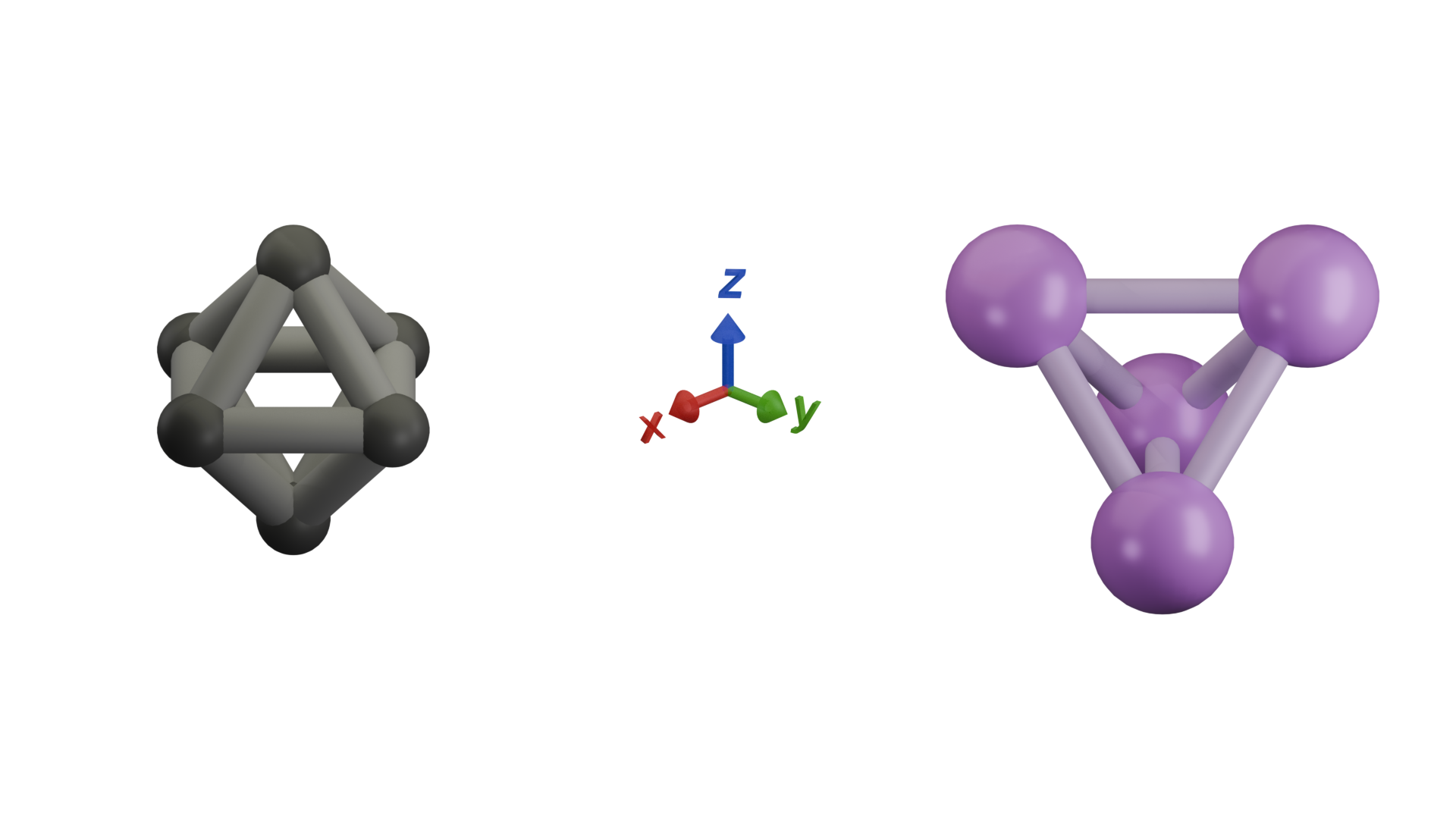}
            \caption{%
              Geometries of octahedral (H\textsubscript{6})\textsuperscript{2+} and H\textsubscript{6} (left), and tetrahedral Li\textsubscript{4} (right).
              Atom colours: \ce{H} --- black, \ce{Li} --- purple.
            }
            \label{fig:structures}
        \end{figure}

        The proposed method was examined on three highly symmetric molecules, namely (H\textsubscript{6})\textsuperscript{2+}, H\textsubscript{6}, and Li\textsubscript{4} (Figure~\ref{fig:structures}). 
        Both (H\textsubscript{6})\textsuperscript{2+} and H\textsubscript{6} have $\mathcal{O}_h$ point-group symmetry, whereas Li\textsubscript{4} has $\mathcal{T}_d$ point-group symmetry.
        The high symmetry of these systems leads to degenerate low-lying electronic states on which the capability of GNOCCSD was benchmarked against existing correlated electronic-structure approaches.
        All calculations were performed in the 6-31G basis set.
        
      \subsection{Symmetry adaptation of symmetry-broken RHF determinants}
        For each molecular system, \textsc{PySCF}\cite{sunPySCFPythonbasedSimulations2018,sunRecentDevelopmentsPySCF2020,sunLibcintEfficientGeneral2015} was employed to identify the lowest RHF solution by a combination of self-consistent-field convergence via the Direct Inversion in the Iterative Subspace algorithm\cite{article:Pulay1980} with a convergence threshold of \num{e-11} and stability analysis (\textit{vide infra}).\cite{seegerSelfconsistentMolecularOrbital1977}
        This ensures that any symmetry breaking found is genuine to the RHF solution and not an artefact of non-convergence.
        \qsymsq{} was then used to form the appropriate symmetry orbit for each RHF solution, from which its symmetry assignment was found and the symmetry-conserved NOCI wavefunctions based on it were computed.
        The procedure outlined in Section~\ref{sssection:NOCI2CAS} and summarised in Figure~\ref{fig:cas-type-construction} was then followed to construct corresponding CAS-type wavefunctions ready for correlation by GNOCCSD.

      \subsection{Correlated calculations}
        Full Configuration Interaction (FCI) calculations were performed using \textsc{PySCF} and 
        Iterative-Configuration Expansion Configuration Interaction (ICE-CI)\cite{chilkuriComparisonManyparticleRepresentations2021,chilkuriComparisonManyparticleRepresentations2021a} calculations were performed using \textsc{ORCA}.\cite{neeseORCAProgramSystem2012,neeseORCAQuantumChemistry2020,neeseSoftwareUpdateORCA2022}
        FCI was used to benchmark state energies obtained from GNOCCSD, with ICE-CI used as a fall-back option should FCI become computationally unfeasible. 
 
        CCSD and CCSD(T) calculations were also performed using \textsc{PySCF} with both RHF and UHF solutions used as reference determinants in our best attempts to target specific ground and excited states with single-reference coupled cluster.\cite{article:Lee2019a}
        Specifically, the lowest RHF and UHF solutions were found by performing a stability analysis\cite{article:Paldus1990} and ensuring that the solution has no internal instabilities.
        The excited RHF solutions were located by enforcing symmetry constraints provided by \textsc{PySCF}.
        
        On the other hand, Equation-of-Motion Coupled Cluster Singles and Doubles (EOM-CCSD) calculations were performed using \textsc{ORCA} with the lowest UHF solution as its reference determinant for each system.
        Since EOM-CCSD is not a spatial-symmetry-adapted method, it is challenging to assign a specific electronic state to each excitation energy found.
        Therefore, EOM-CCSD excitation energies were assigned to electronic states by comparison with benchmark values (FCI or ICE-CI), selecting in each case the EOM-CCSD result with the smallest absolute deviation from the reference.

        GNOCCSD calculations were run with in-house coupled-cluster code \gnocchi{}, and were all converged to a residual norm of \num{e-7}.
        Unless otherwise stated, the redundancy threshold (\textit{cf.} Section~\ref{sec:gnocc}) for selecting a non-null excitation space is \num{e-8}.

\section{Results and Discussion}
    \label{sec:results}
    \subsection{Octahedral (H\textsubscript{6})\textsuperscript{2+}}
      \subsubsection{The reference wavefunctions}

\begin{figure}[h!]
          \centering
          \includegraphics[width=0.6\linewidth]{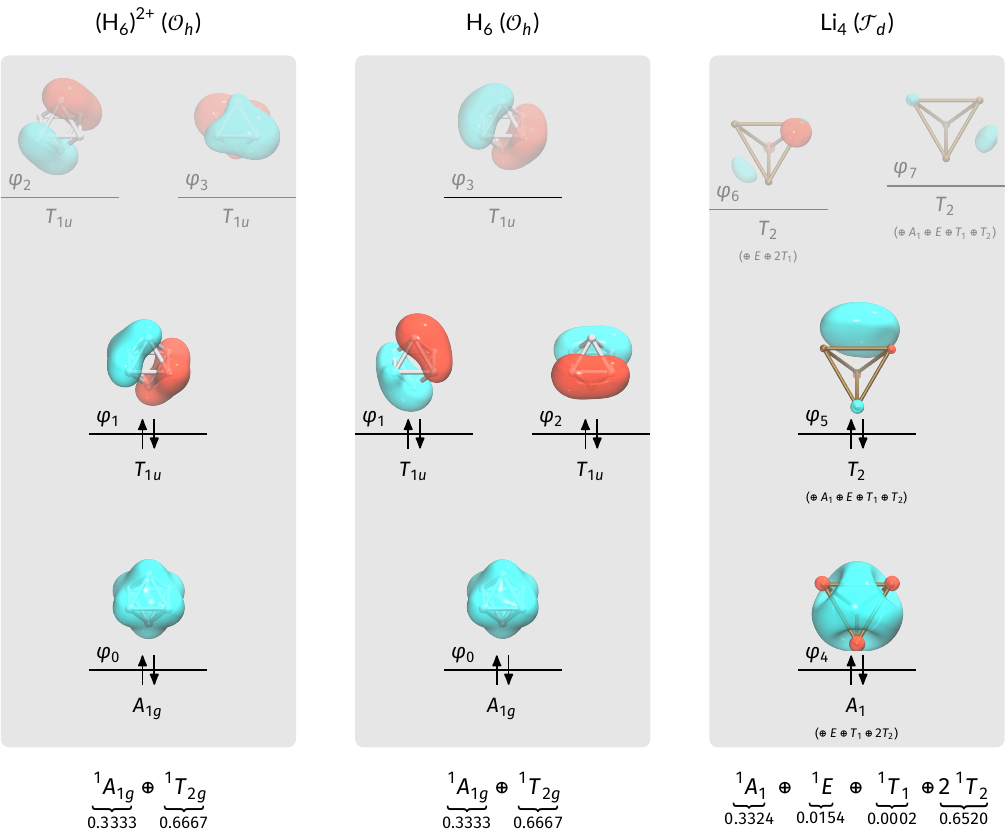}
          \caption{%
            Electronic configurations of the symmetry-broken RHF wavefunctions $\ket{\phi_{\mathrm{RHF}}}$ for octahedral (H\textsubscript{6})\textsuperscript{2+} (left panel), octahedral H\textsubscript{6} (middle panel), and tetrahedral Li\textsubscript{4} (right panel), from which the spatial-symmetry-adapted NOCI wavefunctions employed for GNOCCSD in this work are constructed.
            The configurations are described by the isosurfaces and symmetries of the frontier molecular orbitals in grey boxes.
            The overall symmetries of the RHF wavefunctions are shown below the grey boxes together with their compositions given by $\braket{\phi^{\mathrm{RHF}} | \hat{P}_{\Gamma} | \phi^{\mathrm{RHF}}}$ (\textit{cf.} Equation~\ref{eq:Phi-Gamma_i-proj}) as determined by \qsymsq{}.
            The frontier molecular orbitals in (H\textsubscript{6})\textsuperscript{2+} and H\textsubscript{6} are all symmetry-conserved, whereas those in Li\textsubscript{4} are slightly symmetry-broken; non-dominant components with compositions smaller than $0.01$ are shown at reduced size.
          }
          \label{fig:rhf-states}
        \end{figure}

        The left panel of Figure~\ref{fig:rhf-states} shows the frontier molecular orbitals of the lowest RHF wavefunction $\ket{\phi^{\mathrm{RHF}}}$ that has been located for (H\textsubscript{6})\textsuperscript{2+} and verified with stability analysis.
        Inspection of the symmetry and occupation of these molecular orbitals reveals that $\ket{\phi^{\mathrm{RHF}}}$ is an exact spin eigenfunction with $S = 0$ and corresponds to the configuration $(a_{1g})^2 (t_{1u})^2$.
        As such, it is expected to describe one or more of the terms $\prescript{1}{}{A}_{1g} \oplus \prescript{1}{}{E}_g \oplus \prescript{1}{}{T}_{2g}$.
        A detailed analysis of the symmetry of $\ket{\phi^{\mathrm{RHF}}}$ via \qsymsq{} shows that it is indeed symmetry-broken and constitutes a basis that spans $\prescript{1}{}{A}_{1g} \oplus \prescript{1}{}{T}_{2g}$.
        Consequently, the symmetry adaptation procedure via either NOCI or symmetry projection as described in Section~\ref{sec:symmetry} must yield symmetry-conserved multi-determinantal wavefunctions $\ket{\Phi}$ that have either $\prescript{1}{}{A}_{1g}$ or $\prescript{1}{}{T}_{2g}$ symmetry.
        In fact, we found four optimal NOCI wavefunctions from the orbit  $\mathcal{O}_h \cdot \ket{\phi^{\mathrm{RHF}}}$: $\ket{\Phi_0}, \ket{\Phi_1}, \ket{\Phi_2}$ are triply degenerate with $\prescript{1}{}{T}_{2g}$ symmetry, and $\ket{\Phi_3}$ is non-degenerate with $\prescript{1}{}{A}_{1g}$ symmetry.
        
\begin{figure}[h!]
          \centering
          \includegraphics[width=0.50\linewidth]{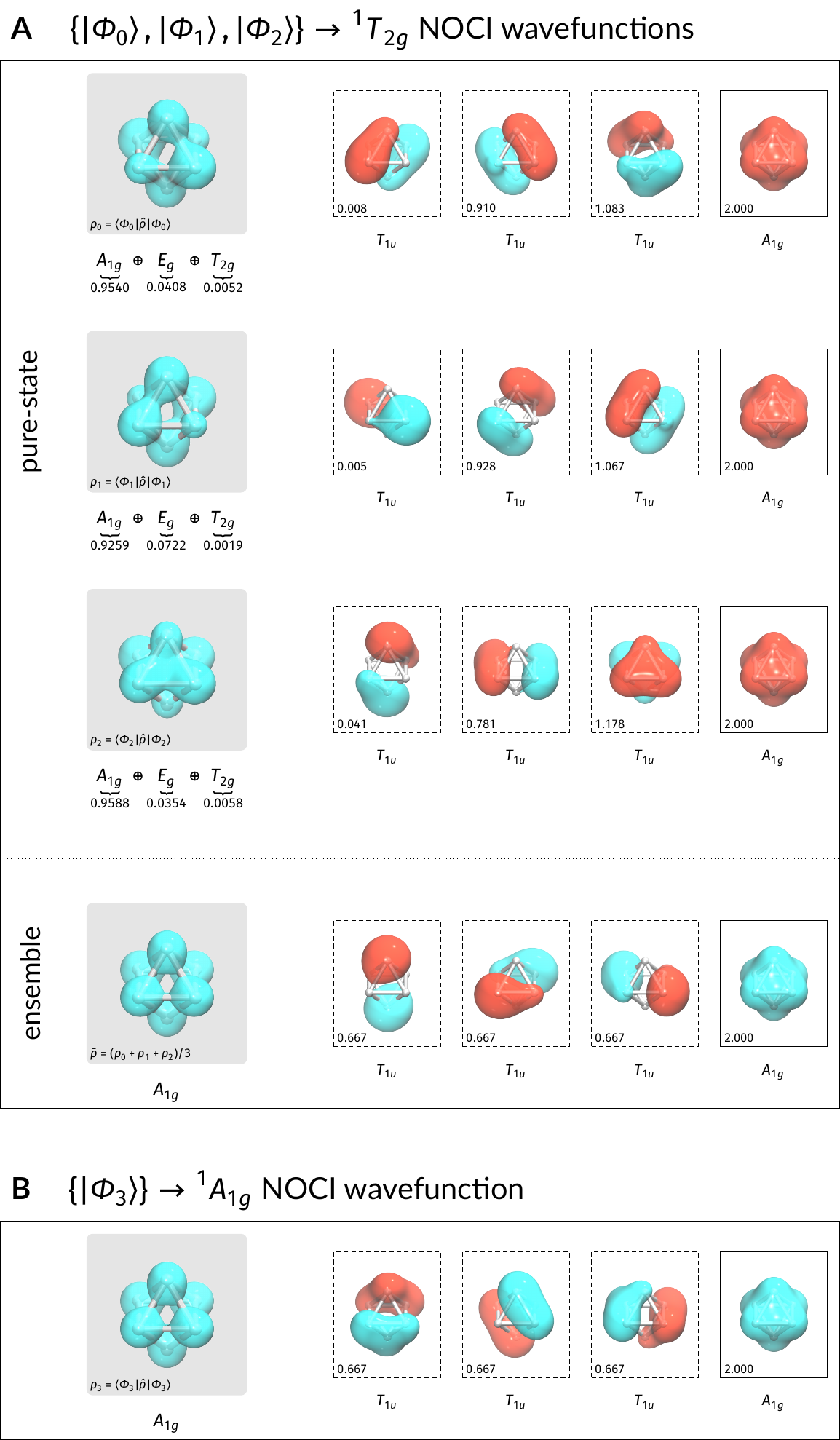}
          \caption{%
            Electron densities $\rho(\mathbf{r})$ and natural orbitals $\{\varphi_k^{\mathrm{NO}}\}$ of the symmetry-conserved NOCI wavefunctions based on the $\prescript{1}{}{A}_{1g} \oplus \prescript{1}{}{T}_{2g}$ symmetry-broken RHF wavefunction for (H\textsubscript{6})\textsuperscript{2+} with $\mathcal{O}_h$ symmetry.
            The isosurface of each electron density is shown at isovalue $\rho(\mathbf{r}) = 0.052 a_0^{-3}$ on a grey background.
            Symmetry-broken densities are also accompanied by their symmetry compositions given by $\braket{\rho | \hat{P}_{\Gamma} | \rho}$ (\textit{cf.} Equation~\ref{eq:Phi-Gamma_i-proj}) as determined by \qsymsq{}.
            The isosurfaces of the corresponding natural orbitals are shown at isovalue $|\varphi(\mathbf{r})| = 0.080 a_0^{-3/2}$ together with their occupation numbers and symmetries.
            Dashed boxes indicate active natural orbitals and solid boxes indicate core natural orbitals.
            \textbf{A}. Triply degenerate $\prescript{1}{}{T}_{2g}$ NOCI wavefunctions.
            The pure-state densities are symmetry-broken, each belonging to a basis spanning the reducible representation $(T_{2g})^2_+ = A_{1g} \oplus E_g \oplus T_{2g}$.
            The equal-weight ensemble density is, on the other hand, totally symmetric.
            \textbf{B}. Non-degenerate $\prescript{1}{}{A}_{1g}$ NOCI wavefunction.
            The pure-state density is already totally symmetric.
          }
          \label{fig:noci-h6-2p}
        \end{figure}

        The electron densities $\rho(\mathbf{r}) = \prescript{1}{}{\rho}(\mathbf{r})$ (Equation~\ref{eq:spintraced-n-rdm}) and natural orbitals $\{\varphi_k^{\mathrm{NO}}\}$ (Figure~\ref{fig:cas-type-construction}, step \textbf{A}) corresponding to the four NOCI wavefunctions are depicted in Figure~\ref{fig:noci-h6-2p}.
        As discussed in Section~\ref{sec:sym-RDMs}, the pure-state densities associated with the degenerate NOCI wavefunctions $\ket{\Phi_0}, \ket{\Phi_1}, \ket{\Phi_2}$ are symmetry-broken, each belonging to a basis spanning the reducible representation $(T_{2g})^2_+ = A_{1g} \oplus E_g \oplus T_{2g}$ (Figure~\ref{fig:noci-h6-2p}A).
        Although the relative compositions of these irreducible components differ amongst the pure-state densities, the dominant contribution from $A_{1g}$ accounts for their nearly totally-symmetric appearance.
        Nevertheless, despite the small magnitude of the non-totally-symmetric components, variations in their composition are sufficient to induce significant differences in the active natural orbitals across the three NOCI wavefunctions, both in their shape and in their occupation numbers.
        These unphysical non-totally-symmetric components can be removed by constructing the totally symmetric ensemble density, either via the projection operator (Equation~\ref{eq:P-n_gamma_i}) or by forming the equal-weight ensemble density (Equation~\ref{eq:equi-ensemble}).
        The corresponding active natural orbitals then exhibit equal occupation numbers consistent with the triply degenerate nature of their $T_{1u}$ symmetry.
        On the other hand, the non-degenerate NOCI wavefunction $\ket{\Phi_3}$ does not suffer from this issue.
        In fact, as shown in Figure~\ref{fig:noci-h6-2p}B, its pure-state density is already totally symmetric, and the corresponding active natural orbitals already exhibit equal occupation numbers expected for their $T_{1u}$ symmetry.
        Each of the four NOCI wavefunctions is then used in steps \textbf{B} and \textbf{C} of Figure~\ref{fig:cas-type-construction} to compute the associated GNOCCSD correlation energy.
        
      \subsubsection{GNOCCSD correlation}

        For (H\textsubscript{6})\textsuperscript{2+} in the 6-31G basis, the conversion from the symmetry-adapted NOCI wavefunctions $\ket{\Phi_M}$ to the corresponding full CAS-type wavefunctions $\ket{\tilde{\Phi}_M}$ is exact for both the $\prescript{1}{}{T}_{2g}$ and $\prescript{1}{}{A}_{1g}$ states.
        Indeed, all four states $\ket{\tilde{\Phi}_M}$ ($M = 0, \ldots, 3$) can be represented exactly within a CAS(3o,2e) framework.
        This can be attributed to the fact that, in each case, the core natural orbital lies entirely within the occupied orbital space of every determinant in the NOCI expansion.
        This observation is supported by the close correspondence in both form and symmetry between the RHF $\varphi_0$ molecular orbital (left panel of Figure~\ref{fig:rhf-states}) and the core natural orbitals shown in Figure~\ref{fig:noci-h6-2p}.
        The exactness of this conversion is further confirmed numerically by the vanishing \emph{CAS-type construction error} given by the energy difference $E_{\ket{\tilde{\Phi}_M}} - E_{\ket{\Phi_M}}$ (Equation~\ref{eq:cas-error}), as shown in Table~\ref{tab:H6_2p_Oh_Energies}.
        
        For comparison, in single-reference CCSD and CCSD(T) calculations, a HF reference must first be chosen, ideally one that shares key characteristics (\textit{e.g.}, symmetry) with the target excited state.\cite{article:Lee2019a,article:Foerster2025}
        In the present system, however, all low-lying UHF and RHF solutions identified are symmetry-broken, leading to ambiguity in the choice of reference and in the subsequent assignment of CCSD and CCSD(T) energies to specific electronic states.
        Fortunately, the availability of Full Configuration Interaction (FCI) benchmark energies provides a means to resolve this ambiguity, allowing the CCSD and CCSD(T) results to be reliably assigned by direct comparison.
        In any case, the $\prescript{1}{}{T}_{2g}$ state is approximated using either the lowest symmetry-broken UHF reference that has $T_{1g} \oplus T_{2g}$ spatial symmetry and $\langle \hat{S}^2 \rangle \approx 1.081$ or the lowest symmetry-broken RHF reference that has $A_{1g} \oplus T_{2g}$ spatial symmetry (which is also the symmetry-broken RHF wavefunction shown in the left panel of Figure~\ref{fig:rhf-states} that is used to construct the NOCI references for our GNOCCSD calculations), whereas the $\prescript{1}{}{A}_{1g}$ state is described using an excited RHF reference that has $A_{1g} \oplus E_g$ spatial symmetry.
        
        In Table~\ref{tab:H6_2p_Oh_Energies}, we report the errors in the state energies computed with GNOCCSD relative to FCI for the $\prescript{1}{}{T}_{2g}$ and $\prescript{1}{}{A}_{1g}$ states, which are the ground and first excited states within the singlet spin manifold, respectively.
        The GNOCCSD results are also benchmarked against CCSD, CCSD(T), and EOM-CCSD, with due consideration of the reference-state ambiguities discussed above.
        While CCSD and CCSD(T) yield reasonably accurate energies for the ground $\prescript{1}{}{T}_{2g}$ state, albeit with some dependence on the chosen reference, they perform poorly for the excited $\prescript{1}{}{A}_{1g}$ state.
        Similarly, EOM-CCSD exhibits large errors for the $\prescript{1}{}{A}_{1g}$ state, comparable to those of CCSD and CCSD(T).
        In contrast, GNOCCSD is able to determine the energies of both states to within chemical accuracy ($\le \SI{1}{\milli\hartree}$).

        Besides numerical performances, it is instructive to analyse the symmetry properties of the reference NOCI states and the correlated coupled cluster wavefunctions.
        The reference NOCI states employed for GNOCCSD preserve both spatial and spin symmetries, and GNOCCSD formalism itself is spin-adapted (notwithstanding some residual spatial symmetry breaking; \textit{c.f.} Section~\ref{sec:residualspatialsymbreaking}).
        Consequently, the GNOCCSD wavefunctions are expected to maintain proper spin symmetry and exhibit reduced spatial symmetry breaking relative to CCSD and CCSD(T) based on symmetry-broken references.

        By contrast, the single-reference calculations employ symmetry-broken HF references.
        In particular, the lowest UHF solution ($T_{1g} \oplus T_{2g}$, $\langle \hat{S}^2 \rangle \approx 1.081$) breaks both spin and spatial symmetries, while the lowest RHF solution ($A_{1g} \oplus T_{2g}$) breaks spatial symmetry.
        The resulting CCSD and CCSD(T) wavefunctions are thus expected to inherit similar symmetry-breaking characteristics, which are known to artificially lower the predicted energies.

        The ability to target ground and excited states of any given symmetry, with comparable accuracy, is encouraging for its use in excited-state chemistry.
        
        \begin{table}[h!]
          \centering
          \renewcommand{\arraystretch}{1.5}
          \caption{%
            (H\textsubscript{6})\textsuperscript{2+}: CAS-type construction error ($E_{\ket{\tilde{\Phi}_M}} - E_{\ket{\Phi_M}}$), FCI state energy, and state energy error relative to FCI for CCSD, CCSD(T), EOM-CCSD, and GNOCCSD (this work). 
            CCSD and CCSD(T) energies for the $\prescript{1}{}{T}_{2g} $ state were approximated by using either the lowest symmetry-broken UHF reference with spatial symmetry $T_{1g} \oplus T_{2g}$ and $\langle \hat{S}^2 \rangle \approx 1.081$ ($\dagger$) or the lowest symmetry-broken RHF reference with spatial symmetry $A_{1g} \oplus T_{2g}$ ($\ddagger$), whereas those for the $\prescript{1}{}{A}_{1g} $ state were found by using an excited RHF reference with symmetry $A_{1g} \oplus E_g$.
          }
          \label{tab:H6_2p_Oh_Energies}
          \begin{tabular}{%
            l%
            S[table-format=+1.3, table-alignment-mode=format, table-number-alignment=right, table-text-alignment=right]%
            S[table-format=+1.6, table-alignment-mode=format, table-number-alignment=right, table-text-alignment=right] |%
            S[table-format=+2.3, table-alignment-mode=format, table-number-alignment=right, table-text-alignment=right]%
            S[table-format=+1.3, table-alignment-mode=format, table-number-alignment=right, table-text-alignment=right]%
            S[table-format=+1.3, table-alignment-mode=format, table-number-alignment=right, table-text-alignment=right]%
            S[table-format=+1.3, table-alignment-mode=format, table-number-alignment=right, table-text-alignment=right]%
          }
            \hline\hline
            \multirow{2}{*}{Term}     & {\multirow{2}{*}{$E_{\ket{\tilde{\Phi}_M}} - E_{\ket{\Phi_M}}$ / $\si{\milli\hartree}$}} & {\multirow{2}{*}{FCI energy / $\si{\hartree}$}} & \multicolumn{4}{c}{State energy error relative to FCI / $\si{\milli\hartree}$} \\
                                      &                                                                  &                                                 & {CCSD}              & {CCSD(T)}           & {EOM-CCSD} & {This work}           \\
            \hline
            $\prescript{1}{}{T}_{2g}$ & 0.000                                                            & -1.833029                                       & 0.281$^{\dagger}$   & 0.156$^{\dagger}$   & {-}        & 0.698                 \\
                                      &                                                                  &                                                 & 3.155$^{\ddagger}$ & 1.251$^{\ddagger}$ &            &                       \\
            $\prescript{1}{}{A}_{1g}$ & 0.000                                                            & -1.723622                                       & 13.975              & 5.059               & 9.061      & 0.531                 \\
            \hline\hline
          \end{tabular}
        \end{table}

    \subsection{Octahedral H\textsubscript{6}}
      \subsubsection{The reference wavefunctions}

        The middle panel of Figure~\ref{fig:rhf-states} depicts the frontier molecular orbitals of the lowest RHF wavefunction $\ket{\phi^{\mathrm{RHF}}}$ identified for H\textsubscript{6} and confirmed by stability analysis.
        The orbital symmetries and occupations indicate that $\ket{\phi^{\mathrm{RHF}}}$ is an exact spin eigenfunction with $S = 0$ and corresponds to the configuration $(a_{1g})^2 (t_{1u})^4$.
        Therefore, $\ket{\phi^{\mathrm{RHF}}}$ is also expected to describe one or more of the terms $\prescript{1}{}{A}_{1g} \oplus \prescript{1}{}{E}_g \oplus \prescript{1}{}{T}_{2g}$, which are identical to those for (H\textsubscript{6})\textsuperscript{2+}, as expected from the equivalence of the $(t_{1u})^2$ and $(t_{1u})^4$ configurations.
        Accordingly, the lowest RHF solution for H\textsubscript{6} breaks symmetry in an analogous manner to (H\textsubscript{6})\textsuperscript{2+} ($\prescript{1}{}{A}_{1g} \oplus \prescript{1}{}{T}_{2g}$).
        The corresponding orbit $\mathcal{O}_h \cdot \ket{\phi^{\mathrm{RHF}}}$ therefore gives rise to four optimal NOCI wavefunctions: a triply degenerate set comprising $\ket{\Phi_0}, \ket{\Phi_1}, \ket{\Phi_2}$ with $\prescript{1}{}{T}_{2g}$ symmetry, and a non-degenerate $\ket{\Phi_3}$ with $\prescript{1}{}{A}_{1g}$ symmetry.
        
      \subsubsection{GNOCCSD correlation}
        The results for H\textsubscript{6} in the 6-31G basis follow similar trends to those observed for (H\textsubscript{6})\textsuperscript{2+}.
        In particular, the conversion from symmetry-adapted NOCI wavefunctions to CAS-type wavefunctions remains exact for both the $\prescript{1}{}{T}_{2g}$ and $\prescript{1}{}{A}_{1g}$ states, with all four NOCI wavefunctions expressible within a CAS(3o,4e) framework.
        Just as for (H\textsubscript{6})\textsuperscript{2+}, single-reference methods yield significantly larger deviations from FCI for the excited $\prescript{1}{}{A}_{1g}$ state energy than for the ground $\prescript{1}{}{T}_{2g}$ state energy.
        By contrast, GNOCCSD based on symmetry-conserved NOCI references still delivers comparable accuracy for both states, once again showcasing its ability to afford a balanced treatment of ground and excited states.
        However, unlike in (H\textsubscript{6})\textsuperscript{2+}, the GNOCCSD errors for H\textsubscript{6} fall outside of chemical accuracy.
        This is due to the addition of two more electrons to the active space in H\textsubscript{6}, which renders triple excitations, and thus three-electron correlation, non-negligible.
        The inclusion of these excitations are required to recover chemical accuracy and will be addressed in future work. 
        
        \begin{table}[h!]
          \centering
          \renewcommand{\arraystretch}{1.5}
          \caption{
            H\textsubscript{6}: CAS-type construction error ($E_{\ket{\tilde{\Phi}_M}} - E_{\ket{\Phi_M}}$), FCI state energy, and state energy error relative to FCI for CCSD, CCSD(T), EOM-CCSD, and GNOCCSD (this work). 
            CCSD and CCSD(T) energies for the $\prescript{1}{}{T}_{2g} $ state were approximated by using either the lowest symmetry-broken UHF reference with spatial symmetry $T_{1g} \oplus T_{2g}$ and $\langle \hat{S}^2 \rangle \approx 1.082$ ($\dagger$) or the lowest symmetry-broken RHF reference with spatial symmetry $A_{1g} \oplus T_{2g}$ ($\ddagger$), whereas those for the $\prescript{1}{}{A}_{1g} $ state were found by using an excited RHF reference with symmetry $A_{1g} \oplus E_g$.
          }
          \label{tab:H6_Oh_Energies}
            \begin{tabular}{%
            l%
            S[table-format=+1.3, table-alignment-mode=format, table-number-alignment=right, table-text-alignment=right]%
            S[table-format=+1.6, table-alignment-mode=format, table-number-alignment=right, table-text-alignment=right] |%
            S[table-format=+2.3, table-alignment-mode=format, table-number-alignment=right, table-text-alignment=right]%
            S[table-format=+1.3, table-alignment-mode=format, table-number-alignment=right, table-text-alignment=right]%
            S[table-format=+1.3, table-alignment-mode=format, table-number-alignment=right, table-text-alignment=right]%
            S[table-format=+1.3, table-alignment-mode=format, table-number-alignment=right, table-text-alignment=right]%
          }
            \hline\hline
            \multirow{2}{*}{Term}     & {\multirow{2}{*}{$E_{\ket{\tilde{\Phi}_M}} - E_{\ket{\Phi_M}}$ / $\si{\milli\hartree}$}} & {\multirow{2}{*}{FCI energy / $\si{\hartree}$}} & \multicolumn{4}{c}{State energy error relative to FCI / $\si{\milli\hartree}$} \\
                                      &                                                                  &                                                 & {CCSD}              & {CCSD(T)}           & {EOM-CCSD} & {This work}           \\
            \hline
            $\prescript{1}{}{T}_{2g}$ & 0.000                                                            & -2.968770                                       & 2.017$^{\dagger}$   & 1.237$^{\dagger}$   & {-}        & 4.462                 \\
                                      &                                                                  &                                                 & 11.081$^{\ddagger}$ & 6.869$^{\ddagger}$ &            &                       \\
            $\prescript{1}{}{A}_{1g}$ & 0.000                                                            & -2.877283                                       & 20.734              & 13.271               & 13.387      & 2.039                 \\
            \hline\hline
          \end{tabular}
        \end{table}

    \subsection{Tetrahedral Li\textsubscript{4}}
      \subsubsection{The reference wavefunctions}
        The right panel of Figure~\ref{fig:rhf-states} shows the frontier molecular orbitals of the RHF wavefunction $\ket{\phi^{\mathrm{RHF}}}$ located for Li\textsubscript{4} that was used in the GNOCCSD calculation.
        The symmetry and occupation of these molecular orbitals indicate that $\ket{\phi^{\mathrm{RHF}}}$ is an exact spin eigenfunction with $S = 0$ and corresponds to the configuration $(a_{1})^2 (t_{2})^2$ \textit{approximately} (since each molecular orbital is also symmetry-broken where the major component shown is contaminated with other, smaller, components).
        The RHF wavefunction $\ket{\phi^{\mathrm{RHF}}}$ is thus expected to describe one or more of the terms $\prescript{1}{}{A}_{1} \oplus \prescript{1}{}{E} \oplus \prescript{1}{}{T}_{2}$.
        In fact, \qsymsq{} shows that $\ket{\phi^{\mathrm{RHF}}}$ breaks symmetry as $\prescript{1}{}{A}_{1} \oplus \prescript{1}{}{E} \oplus \prescript{1}{}{T}_{1} \oplus 2\prescript{1}{}{T}_{2}$ where, as it turns out, $\prescript{1}{}{T}_{1}$ and one of the two occurrences of $\prescript{1}{}{T}_{2}$ are only present as negligible contaminations and will thus be ignored.
        Consequently, the orbit $\mathcal{O}_h \cdot \ket{\phi^{\mathrm{RHF}}}$ gives rise to twelve optimal NOCI wavefunctions, but only six of which are low enough in energy to be of physical significance; the remaining six correspond to the contaminations $\prescript{1}{}{T}_{1}$ and $\prescript{1}{}{T}_{2}$ and are too high in energy. 
        The six relevant NOCI wavefunctions comprise $\ket{\Phi_0}, \ket{\Phi_1}, \ket{\Phi_2}$ which are triply degenerate with $\prescript{1}{}{T}_{2}$ symmetry, $\ket{\Phi_3}, \ket{\Phi_4}$ which are doubly degenerate with $\prescript{1}{}{E}$ symmetry, and $\ket{\Phi_5}$ which is non-degenerate with $\prescript{1}{}{A}_{1}$ symmetry.

        As it turns out, this RHF wavefunction is \textit{not} the lowest RHF solution that exists for Li\textsubscript{4} in 6-31G; an alternative solution with a lower energy exists that transforms as $\prescript{1}{}{A}_{1} \oplus \prescript{1}{}{E} \oplus \prescript{1}{}{T}_{2}$ and therefore does not contain the aforementioned $\prescript{1}{}{T}_{1}$ and $\prescript{1}{}{T}_{2}$ contaminations.
        However, GNOCCSD was found to encounter convergence issues with NOCI wavefunctions based on this alternative RHF solution.
        The reasons for this unexpected behaviour will be investigated in a future study; for this work, though, we shall only focus on GNOCCSD results for NOCI wavefunctions based on the higher-energy and slightly contaminated $\ket{\phi^{\mathrm{RHF}}}$.

      \subsubsection{GNOCCSD correlation}

        For Li\textsubscript{4} in 6-31G, the conversion from symmetry-adapted NOCI wavefunctions to CAS-type wavefunctions for the ground and excited states is now non-exact, with CAS-type construction errors of up to \SI{4.821}{\milli\hartree} for the $\prescript{1}{}{E}$ state, as shown in Table~\ref{tab:Li4_Td_Energies}. 
        This is because strict equalities no longer hold in Equation~\ref{eq:core-spans}: the space spanned by the core natural orbitals is not identical to the core orbital space of every constituent determinant in the NOCI wavefunctions.
        To benchmark GNOCCSD and the other coupled-cluster calculations, we used the ICE-CI method\cite{chilkuriComparisonManyparticleRepresentations2021,chilkuriComparisonManyparticleRepresentations2021a} based on the lowest symmetry-broken RHF solution (\textit{i.e.}, the alternative RHF solution that transforms as $\prescript{1}{}{A}_{1} \oplus \prescript{1}{}{E} \oplus \prescript{1}{}{T}_{2}$ symmetry as described above), with the lowest-energy occupied molecular orbital frozen.
        
        In a similar manner to the previous two systems, the CCSD and CCSD(T) energies for the ground $\prescript{1}{}{T}_{2}$ state were computed based on the lowest-lying UHF reference with symmetry $A_1 \oplus A_2 \oplus 2E \oplus T_{1} \oplus T_{2}$ and $\langle \hat{S}^2 \rangle \approx 1.365$.
        Similarly, the CCSD and CCSD(T) energies for the first-excited $\prescript{1}{}{E}$ state were approximated based on an excited RHF reference with symmetry $A_1 \oplus E$.
        It is clear that should one wish to target multiple excited states using single-reference coupled cluster, one would require multiple unique HF solutions.
        However, we have not attempted to find higher-energy HF solutions and hence restricted to the lowest-UHF and a specific RHF references.
        We therefore could only access the ground $\prescript{1}{}{T}_{2}$ and first-excited $\prescript{1}{}{E}$ states via CCSD and CCSD(T), but not the second-excited state $\prescript{1}{}{A}_{1}$.

        Table~\ref{tab:Li4_Td_Energies} shows that CCSD and CCSD(T) underestimate the energy of the $\prescript{1}{}{T}_{2}$ state relative to the ICE-CI benchmark, while overestimating the energy of the $\prescript{1}{}{E}$ state. 
        This leads to an overestimation of the energy gap between the ground and the first-excited states. 
        On the other hand, EOM-CC yields approximations for the $\prescript{1}{}{E}$ and $\prescript{1}{}{A}_{1}$ states that have errors with opposite signs, which in turn results in an underestimation of the gap between these two states.

        Tetrahedral Li\textsubscript{4} presents an example of a challenging system for which GNOCCSD struggles with convergence. 
        This is mainly because there are certain internally-contracted excitations in the GNOCCSD calculations that give rise to numerical instabilities during the redundancy handling procedure (\textit{cf.} Section~\ref{sec:redundancy}).\cite{leeSpinfreeGeneralizedNormal2026,evangelistaOrbitalinvariantInternallyContracted2011,hanauerPilotApplicationsInternally2011} 
        Therefore, we employed a custom method to trim the excitation space to improve convergence for GNOCCSD calculations for this system, albeit at a minor loss in accuracy, as detailed in Appendix~\ref{appendix:prune}.
        Unlike the single-reference approaches, GNOCCSD consistently delivers energies above the ICE-CI benchmark.
        In fact, this trend has been observed for all three systems that we have investigated thus far, and is particularly noteworthy given the fact that GNOCCSD itself is not a variational method.
        The reasons for this behaviour, as well as the possible ways to improve the convergence of GNOCCSD for similar challenging systems, will be examined in a future study.
        
        \begin{table}[h!]
          \centering
          \renewcommand{\arraystretch}{1.5}
          \caption{
            Li\textsubscript{4}: CAS-type construction error ($E_{\ket{\tilde{\Phi}_M}} - E_{\ket{\Phi_M}}$), ICE-CI state energy, and state energy error relative to ICE-CI for CCSD, CCSD(T), EOM-CCSD, and GNOCCSD (this work). 
            CCSD and CCSD(T) energies for the $\prescript{1}{}{T}_{2} $ state were approximated by using the lowest symmetry-broken UHF reference with spatial symmetry $A_1 \oplus A_2 \oplus 2E \oplus T_{1} \oplus T_{2}$ and $\langle \hat{S}^2 \rangle \approx 1.365$, whereas those for the $\prescript{1}{}{E}$ state were found by using an excited RHF reference with spatial symmetry $A_1 \oplus E$.
            No suitable HF references could be found such that the excited $\prescript{1}{}{A}_1$ state could be targeted with CCSD and CCSD(T).
          }
          \label{tab:Li4_Td_Energies}
          \begin{tabular}{%
            l%
            S[table-format=+1.3, table-alignment-mode=format, table-number-alignment=right, table-text-alignment=right]%
            S[table-format=+1.6, table-alignment-mode=format, table-number-alignment=right, table-text-alignment=right] |%
            S[table-format=+2.3, table-alignment-mode=format, table-number-alignment=right, table-text-alignment=right]%
            S[table-format=+1.3, table-alignment-mode=format, table-number-alignment=right, table-text-alignment=right]%
            S[table-format=+1.3, table-alignment-mode=format, table-number-alignment=right, table-text-alignment=right]%
            S[table-format=+1.3, table-alignment-mode=format, table-number-alignment=right, table-text-alignment=right]%
          }
            \hline\hline
            \multirow{2}{*}{Term}     & {\multirow{2}{*}{$E_{\ket{\tilde{\Phi}_M}} - E_{\ket{\Phi_M}}$ / $\si{\milli\hartree}$}} & {\multirow{2}{*}{ICE-CI energy / $\si{\hartree}$}} & \multicolumn{4}{c}{State energy error relative to ICE-CI / $\si{\milli\hartree}$} \\
                                      &                                                                  &                                                 & {CCSD}              & {CCSD(T)}           & {EOM-CCSD} & {This work}           \\
            \hline
            $\prescript{1}{}{T}_{2}$ &  4.468  &  -29.750306   & -4.081   & -6.428  & {-}         & 4.070\\
            $\prescript{1}{}{E}$       &  4.821  &  -29.745287   & 6.154    & 1.745   & 3.239   & 0.444\\
            $\prescript{1}{}{A}_{1}$  &  0.829   &  -29.727565   & {-}          & {-}         & -5.404  & 2.385\\
            \hline\hline
          \end{tabular}
        \end{table}

\section{Conclusions}
  We have developed a method for obtaining correlated wavefunctions for highly symmetric molecules within the framework of coupled cluster theory.
  Owing to their non-Abelian point-group symmetry, such systems often possess degenerate electronic states that are challenging for single-reference methods.  
  Our approach targets these states by making use of symmetry orbits formed from symmetry-broken RHF determinants via the symbolic framework of \qsymsq{}, generating spatial-symmetry-adapted multi-determinantal wavefunctions. 
  The multi-determinantal nature of these wavefunctions inherently requires a multi-reference treatment to incorporate further correlation effects.
  To this end, we first approximate general multi-determinantal wavefunctions as CAS-type wavefunctions based on considerations of natural orbitals that can be obtained from appropriately symmetrised 1-RDMs.
  Dynamical correlation effects are subsequently incorporated into the resulting CAS-type wavefunctions using the GNOCCSD method.

  This approach enables an accurate and state-specific correlation treatment of both degenerate and non-degenerate electronic states.
  In particular, the spatial symmetry adaptation of the multi-determinantal reference ensures that the resulting correlated wavefunction targets electronic states belonging to a well-defined irreducible representation of the underlying symmetry group.
  Furthermore, numerical results for representative highly symmetric systems, namely octahedral (H\textsubscript{6})\textsuperscript{2+}, octahedral H\textsubscript{6}, and tetrahedral Li\textsubscript{4}, demonstrate that the method yields accurate correlation energies and provides a balanced description of both ground and excited electronic states when compared to widely adopted single-reference coupled-cluster methods. 
  These findings highlight the promise of the approach for applications in which reliable excitation energies between electronic states are essential.
  These results also suggest that symGNOCC may provide a useful foundation for constructing potential energy surfaces of electronic states exhibiting degeneracies, provided that the underlying multi-determinantal references and the corresponding non-redundant excitation spaces can be chosen smoothly and robustly along molecular coordinates.

  However, throughout this article, the term \textit{excited state} refers to any electronic state other than the overall ground state. 
  This definition therefore includes states that are the lowest-energy (ground) states within their respective symmetry sectors.
  In principle, both ground and excited states within a given symmetry sector can be computed, as previously demonstrated by two of the authors.\cite{leeSpinfreeGeneralizedNormal2026} 
  In practice, however, the convergence of excited states within a given symmetry sector using symGNOCC can be challenging.
  This behaviour appears to arise from the sensitivity of the method to both the optimisation algorithm and the choice of excitation space.
  Improving the robustness of the convergence procedure will therefore be an important focus of future work.
  
  Beyond improving convergence, the accuracy of the method is expected to benefit from the inclusion of higher-order correlation effects.
  One possibility would be to include the effects of triple excitations perturbatively, which may allow for the determination of correlation energies to within chemical accuracy.
  We note, however, that the overall GNOCCSD wavefunction may still exhibit symmetry breaking because the exponential cluster operator in GNOCCSD is not guaranteed to transform as the totally symmetric irreducible representation. 
  Addressing this limitation constitutes an important direction for future work.

\section*{Data availability}
  Data for this article, including calculation inputs and raw outputs, are available at Zenodo at \href{https://doi.org/10.5281/zenodo.19637760}{https://doi.org/10.5281/zenodo.19637760}.
  \noindent The code for \qsymsq{} can be found at \href{https://gitlab.com/bangconghuynh/qsym2}{https://gitlab.com/bangconghuynh/qsym2}; the version of the code employed for this study is \texttt{v0.13.1}.

  % A data availability statement (DAS) is required to be submitted alongside all articles. Please read our \href{https://www.rsc.org/journals-books-databases/author-and-reviewer-hub/authors-information/prepare-and-format/data-sharing/#dataavailabilitystatements}{full guidance on data availability statements} for more details and examples of suitable statements you can use.

%%%ABSTRACT%%%%

\section*{Author contributions}
  \textbf{Nicholas Lee}: Conceptualisation (equal); Data curation (lead); Formal analysis (equal); Investigation (lead); Validation (equal); Writing -- original draft (lead); Writing -- review \& editing (equal).
  \textbf{David P. Tew}: Investigation (supporting); Writing -- original draft (supporting); Writing -- review \& editing (equal).
  \textbf{Bang C. Huynh}: Conceptualisation (equal); Data curation (equal); Formal analysis (equal); Investigation (equal); Validation (lead); Visualisation (lead); Writing -- original draft (equal); Writing -- review \& editing (lead).
  
  % We strongly encourage authors to include author contributions and recommend using \href{https://casrai.org/credit/}{CRediT} for standardised contribution descriptions. Please refer to our general \href{https://www.rsc.org/journals-books-databases/journal-authors-reviewers/author-responsibilities/}{author guidelines} for more information about authorship.

\section*{Conflicts of interest}
  There are no conflicts to declare.

\section*{Acknowledgements}
  BCH thanks New College, University of Oxford for funding from the Oglander Fellowship.

%%%END OF MAIN TEXT%%%

%  For footnotes in the main text of the article please number the footnotes to avoid duplicate symbols. e.g.  \footnote[num]{your text} the corresponding author \ast counts as footnote 1, ESI as footnote 2, e.g. if there is no ESI, please start at [num]=[2], if ESI is cited in the title please start at [num]=[3] etc. Please also cite the ESI within the main body of the text using \dag.

% The \balance command can be used to balance the columns on the final page if desired. It should be placed anywhere within the first column of the last page.

% \balance

% If notes are included in your references you can change the title from 'References' to 'Notes and references' using the following command:
% \renewcommand\refname{Notes and references}

\appendix
\section{Pruning of the GNOCCSD excitation space}
\label{appendix:prune}

  In this Appendix, we detail the method for pruning the GNOCCSD excitation space.
  We recall that the canonical orthogonalisation matrix $\mathbf{X}$ from Equation~\ref{eqn:Redundancy} is used to transform a given excitation basis into an equivalent linearly independent basis.
  In practice, the use of $\mathbf{X}$ may be numerically unstable if $\mathbf{X}$ contains large singular values. 
  This is because one of the steps in the GNOCCSD procedure involves transforming the residual vector $R_{\mu} = \bra{\Phi} \hat{\tau}_{\mu}^{\dagger} \hat{H} \{ e^{\hat{T}} \} \ket{\Phi}_{\mathrm{c}}$ into $\tilde{R}_{\nu}$ such that
  \begin{equation}
    \tilde{R}_{\nu} = \sum_{\mu i} X_{\nu i} X_{i \mu}^{\dagger}  R_{\mu}.
  \end{equation}
  Therefore, we will identify and remove the components in $\mathbf{X}$ that give rise to large singular values. 
  This can be done via SVD:
  \begin{equation}
     \mathbf{X} = \mathbf{U} \boldsymbol{\Sigma} \mathbf{V}^{\dagger},
  \end{equation}
  where $\mathbf{U}$ and $\mathbf{V}^{\dagger}$ are both unitary matrices and $\boldsymbol{\Sigma}$ is a rectangular diagonal matrix containing the singular values.
  By construction, $\mathbf{X}$ is a rectangular matrix of dimensions $N_{\mathrm{exc}} \times N_{\mathrm{indep}}$ with non-zero singular values (Equation~\ref{eqn:CanonicalOrth}). 
  We will form a rectangular matrix $\tilde{\mathbf{V}}$ of dimensions $N_{\mathrm{indep}} \times N_{\mathrm{small}}$ from the $N_{\mathrm{small}}$ columns of $\mathbf{V}$ corresponding to singular values in $\boldsymbol{\Sigma}$ smaller than a given threshold $\eta_{\mathrm{small}}$.
  The modified canonical orthogonalisation matrix used is then given by
  \begin{equation}
    \tilde{\mathbf{X}} =  \mathbf{X} \tilde{\mathbf{V}}.
  \end{equation}
  We will use $\tilde{\mathbf{X}}$ in place of $\mathbf{X}$ in the GNOCCSD calculations where this pruning procedure is required.
  For the specific example of Li\textsubscript{4}, we used the numerical threshold $\eta_{\mathrm{small}} = 5.0$.

%%%REFERENCES%%%

\setlength{\bibsep}{3pt plus 0.7ex}
\bibliography{bib/sym-gnocc} %You need to replace "rsc" on this line with the name of your .bib file
\bibliographystyle{rsc} %the RSC's .bst file

\end{document}